\documentclass[aps,twocolumn,preprintnumbers,floats,nofootinbib]{revtex4}\def\@cite#1#2{\textsuperscript{[{#1\if@tempswa , #2\fi}]}}

\usepackage{graphicx}
\usepackage{amsmath}
\usepackage{bm}
\usepackage{amsfonts}
\usepackage{amssymb}
\usepackage{color}
\usepackage{subfigure}
\usepackage{epsfig}
\usepackage{morefloats}
\usepackage{multirow}
\usepackage{graphicx,booktabs}
\usepackage{mathrsfs}
\usepackage{txfonts}
\usepackage{footnote}
\usepackage{pifont}
\usepackage{indentfirst}
\usepackage{graphicx,booktabs}
\usepackage{longtable,lscape}
\usepackage[figuresright]{rotating}

\usepackage[normalem]{ulem}

\usepackage[colorlinks, citecolor=blue,anchorcolor=red,menucolor=red, linkcolor=red,filecolor=red,urlcolor=blue,frenchlinks=red]{hyperref}
\begin{document}

\title{Bottomonia in an unquenched quark model}

\author{Ru-Hui Ni$^{1}$~\footnote {E-mail: niruhui@ucas.ac.cn},
Qian Deng$^{1}$,
Jia-Jun Wu$^{2,4}$~\footnote{E-mail: wujiajun@ucas.ac.cn},
and Xian-Hui Zhong$^{1,3}$~\footnote {E-mail: zhongxh@hunnu.edu.cn}}

\affiliation{ 1) Department of Physics, Hunan Normal University, and Key Laboratory of Low-Dimensional Quantum Structures and Quantum Control of Ministry of Education, Changsha 410081, China }

\affiliation{ 2) School of Physical Sciences, University of Chinese Academy of Sciences, Beijing 100049, China }

\affiliation{ 3)  Synergetic Innovation Center for Quantum Effects and Applications (SICQEA),
Hunan Normal University, Changsha 410081, China}

\affiliation{ 4) Southern Center for Nuclear-Science Theory (SCNT), Institute of Modern Physics, Chinese Academy of Sciences, Huizhou 516000, Guangdong Province, China }


\begin{abstract}
The bottomonium spectrum is systematically studied within an unquenched quark model.
Based on a good description of both the masses and widths for the well-established states,
we further give predictions for the higher $S$-, $P$-, and $D$-wave bottomonium states up to a mass region of $\sim 11.3$ GeV.
For the vector states, the $S$-$D$ mixing and dielectron decays are studied.
Additionally, to understand the role of the higher vector resonances in the $e^{+}e^{-}$ annihilation reaction, we evaluate the cross section by combining our quark model predictions for the mass, dielectron and strong decay properties.
It is found that (i) The mass shifts of the high $b\bar{b}$ states due to the coupled-channel effects are the order of a few tens MeV, most of the high-lying resonances contain significant non-$b\bar{b}$ components.
(ii) The $\Upsilon_1(3D,5D,6D)$ states significantly mix with $\Upsilon(4S,6S,7S)$, respectively, which is mainly induced by the intermediate hadronic loops.
(iii) The non-$b\bar{b}$ components will lead a significant suppression for the dielectron decay widths of some vector resonances.
(iv) The threshold effects of open-bottom meson pairs can cause rich bump structures in the cross section of $e^{+}e^{-}\to b\bar{b}$.
Our model shows that the $\Upsilon(10753)$ may arise from threshold effects due to the strong coupling between $\Upsilon(4S)$ and $\bar{B}^*B^*$.
\end{abstract}


\maketitle

\section{Introduction}

The bottomonium ($b\bar{b}$) spectrum was initially established with the discovery of three vector states $\Upsilon(1S)$, $\Upsilon(2S)$, and $\Upsilon(3S)$ by the E288 Collaboration at Fermilab in 1977 through the proton-nucleus collisions~\cite{Herb:1977ek, Innes:1977ae}.
Subsequently, these vector states can be produced by the $e^{+}e^{-}$ collisions at the Cornell Electron Storage Ring (CESR).
In the early 1980s, the low-lying $P$-wave states $\chi_{bJ}(1P)$~\cite{Klopfenstein:1983nx, Pauss:1983pa} and $\chi_{bJ}(2P)$~\cite{Han:1982zk, Eigen:1982zm} ($J = 0, 1, 2$) were established by the radiative decay processes of $\Upsilon(2S)$ and $\Upsilon(3S)$, respectively.
Since then, more and more bottomonium states have been discovered below the $\bar{B}B$ threshold, including $\eta_b(1S,2S)$, $h_b(1P,2P)$, $\chi_{b1,2}(3P)$, and $\Upsilon_2(1D)$ as listed in the Review of Particle Physics by the Particle Data Group (PDG)~\cite{ParticleDataGroup:2024cfk}.
However, above the $\bar{B}B$ threshold, until now only four vector resonances
$\Upsilon(10580)$, $\Upsilon(10750)$, $\Upsilon(10860)$, and $\Upsilon(11020)$ have been observed by the $e^{+}e^{-}$ collision experiments carried out at CESR, CLEO, BaBar and Belle~\cite{Lovelock:1985nb, CLEO:1984vfn, CLEO:1987iba, BaBar:2004rrm, BaBar:2008cmq,Belle:2019cbt,Belle-II:2022xdi}.

Theoretically, the bottomonium spectrum has been studied by using many approaches and models, such as lattice QCD (LQCD)~\cite{Meinel:2009rd, Burch:2009az, HPQCD:2011qwj}, QCD sum rules~\cite{Wang:2012gj, Azizi:2017izn}, Bethe-Salpeter equation~\cite{Fischer:2014cfa}, light-front quark model~\cite{Choi:2007se, Ke:2010vn, Ke:2010tk}, relativistic flux tube model~\cite{Chen:2019uzm}, various potential models~\cite{Eichten:1980ce,Eichten:1994gt,Godfrey:1985xj,Godfrey:2015dia, Ebert:2011jc,Li:2009nr,Segovia:2016xqb,Wei-Zhao:2013sta,Akbar:2015evy, A:2023bxv,Asghar:2023fvk,Zhao:2023hxc,Kher:2022gbz,Molina:2020zao, Wang:2018rjg,Soni:2017wvy}, and so on.
All of the observed $b\bar{b}$ states below the $\bar{B}B$ threshold together with the $\Upsilon(10580)$ slightly above the $\bar{B}B$ threshold can be well described in theory.
However, for the resonances far above the $\bar{B}B$ threshold, $\Upsilon(10750)$, $\Upsilon(10860)$, and $\Upsilon(11020)$, there still exist some controversies about the nature.
For example, the $\Upsilon(10860)$ and $\Upsilon(11020)$ are commonly explained as the $\Upsilon(5S)$ and $\Upsilon(6S)$, respectively, however, they are suggested as the $b\bar{b}$ $5S-4D$ mixed states in a recent work~\cite{Zhao:2023hxc} due to a congruent matching for both masses and leptonic widths.
For the $\Upsilon(10753)$, one difficultly explains it as any conventional $S$-wave or $D$-wave $b\bar{b}$ states with $J^{PC}=1^{--}$~\cite{Ferretti:2013vua, Godfrey:2015dia, Segovia:2016xqb, Deng:2016ktl, Li:2019qsg, Chen:2019uzm, Wang:2018rjg}.
Thus, many explanations for the $\Upsilon(10753)$, such as a mixed state via $\Upsilon(4S)$-$\Upsilon_1(3D)$ or $\Upsilon(5S)$-$\Upsilon_1(4D)$ mixing~\cite{Bai:2022cfz, Li:2021jjt, Li:2019qsg, Li:2022leg}, a hybrid~\cite{Brambilla:2019esw, TarrusCastella:2021pld}, a tetraquark state~\cite{Ali:2019okl, Bicudo:2020qhp, Bicudo:2022ihz, Wang:2019veq}, have been raised in the literature.

The puzzle surrounding these high-lying states $\Upsilon(10750)$, $\Upsilon(10860)$, and $\Upsilon(11020)$ may mainly arising from the neglecting the ``unquenched coupled-channel effects'' from the open-bottom meson pairs within in the conventional quark model.
These effects can play a crucial role in the excited $b\bar{b}$ states above the $\bar{B}B$ threshold.
In order to include the coupled-channel effects on the mass spectrum, some screened potential models have been proposed in the literature~\cite{Li:2009nr, Gonzalez:2014nka, Vijande:2004he, Segovia:2016xqb, Deng:2016ktl, Wang:2018rjg}, where the widely used linear confinement potential is replaced by a screened form.
However, a recent study on the charmonium spectrum~\cite{Deng:2023mza} has revealed that the mass shifts predicted by a detailed treatment of coupled-channel effects do not follow the typical increasing trend from lower to higher states as often observed in screened potential models.
A key aspect of the coupled-channel dynamics is the involvement of multiple channels with varying thresholds, which are unevenly spaced relative to the mass of the bare valence state.
Additionally, states with different quantum numbers couple to distinct channels, each with different coupling strengths.
As a result, the mass shifts for various bare states induced by these effects are generally unordered, posing a challenge for simple screened potential models to accurately capture these complexities.
Thus, a systematically study of the bottomonium spectrum by seriously including the coupled-channel effects within an unquenched quark model is necessary.

The unquenched coupled-channel effects on hadron spectrum have been discussed for over four decades.
Several pioneering works can be found in Refs.~\cite{Eichten:1978tg, Eichten:1979ms,Heikkila:1983wd, Ono:1983rd, Ono:1985jt}.
In the past decade, the coupled-channel effects on the bottomonium spectrum have been also widely discussed in the literature~\cite{Liu:2011yp, Ferretti:2013vua, Ferretti:2018tco, Lu:2016mbb, Lu:2017hma, Anwar:2018yqm, Bruschini:2021sjh, Bruschini:2021ckr, Husken:2022yik}.
It should be emphasized that in these studies the coupled-channel effects mostly focus on the contributions from the ground meson pairs, $\bar{B}B$, $\bar{B}B^*$, $\bar{B}^*B^*$, $\bar{B}_sB_s$, $\bar{B}_sB_s^*$, and $\bar{B}_s^*B_s^*$~\footnote{For convenience, throughout the paper we adopt abbreviations $\bar{B}B^*$, $\bar{B}_sB_s^*$, and etc. to stand for the $\bar{B}B^*$+c.c., $\bar{B}_sB_s^*$+c.c., and etc., respectively.}, while the high-lying channels involving excited bottom mesons are neglected.
In this work, we further study the bottomonium spectrum within an unquenched quark model, which has achieved a good description of the both the heavy-light meson and charmonium spectrum, recently~\cite{Ni:2023lvx,Deng:2023mza}.
In this unquenched quark model, all of the Okubo-Zweig-Iizuka (OZI)-allowed channels with mass thresholds below the bare states, as well as nearby virtual channels, are included, while the contributions from the channels lying far above the bare states are subtracted from
the dispersion relation by redefining the bare mass with the once-subtracted method~\cite{Pennington:2007xr}.
Furthermore, a factor is adopted to suppress the unphysical contributions of the coupled-channel effects in the high momentum region as done in the literature~\cite{Morel:2002vk, Silvestre-Brac:1991qqx, Zhong:2022cjx, Ortega:2016mms, Ortega:2016pgg, Yang:2022vdb, Yang:2021tvc, Ni:2021pce, Hao:2022vwt, Ni:2023lvx, Deng:2023mza}.

On the other hand, the dielectron decay widths of vector $b\bar{b}$ states are crucial for understanding their production rates in the $e^{+}e^{-} \to \text{hadron}$ process and for revealing their intrinsic nature.
Several studies~\cite{Badalian:2008ik, Badalian:2009bu, Gonzalez:2003gx, Li:2009nr, Godfrey:2015dia, Segovia:2016xqb, Wang:2018rjg, Zhao:2023hxc} have analyzed the dielectron widths of vector $b\bar{b}$ states, many of which consider the $S$-$D$ mixing effect, a phenomenon that significantly influences the dielectron decay width.
However, these studies often overlook the underlying dynamic mechanisms responsible for $S$-$D$ mixing, with the mixing angle typically introduced arbitrarily.
The generation of this mixing angle, induced by intermediate mesonic loops, has been discussed by T\"{o}rnqvist \emph{et al.}~\cite{Heikkila:1983wd, Ono:1983rd} and Lu \emph{et al.}~\cite{Lu:2016mbb}.
However, their analyses are primarily focused on ground-state $B_{(s)}$ mesons, considering $\bar{B}_{(s)}^{(*)}B_{(s)}^{(*)}$ loops.
Due to the limited number of coupling channels considered, these studies are restricted in scope and do not fully explore these effects for vector $b\bar{b}$ states.

Therefore, a comprehensive and systematic investigation of the mass spectra, strong decays, and dielectron decay properties of bottomonium, within a unified framework that incorporates unquenched coupled-channel effects, is currently essential.
The main purposes of this work are as follows:
(i) To obtain a more comprehensive understanding of the $b\bar{b}$ spectrum within the unquenched quark model.
The strong decay properties, the mass shifts, and the $b\bar{b}$-core components
for the high-lying states are systematically evaluated in a unified unquenched framework.
(ii) To better understand the $S$-$D$ mixing dynamic mechanisms for the vector states.
The $S$-$D$ mixing induced by the tensor force and the intermediate hadronic loops are studied, and the mixing angles are evaluated by using these two different mechanisms.
(iii) To deepen our understanding of the dielectron decay properties of vector states in the unquenched framework.
The dielectron decay widths are evaluated by considering both the $S$-$D$ mixing effects and the corrections from the non-$b\bar{b}$ components due to the coupled-channel effects.
(iv) To explore the roles of the high-lying vector resonances and the threshold effects of the open-bottom meson pairs in the total cross section of the $e^+e^-\to b\bar{b}$ reaction.
By combining the dielectron decay widths and spectral density functions derived within the unquenched framework, the cross section for $e^+e^-\to b\bar{b}$ is estimated.
Then, the roles of the vector resonances and threshold effects are shown by the line shape of the cross section.

The paper is organized as follows. In Sec.~\ref{Fram}, we give an introduction of the framework of the unquenched quark model including coupled-channel effects.
the mass spectrum of the $S$-, $P$-, and $D$-wave bottomonium states up to a mass region of $\sim 11.3$ GeV is presented, and related issues are discussed.
In Sec.~\ref{DDiscussion}, the two-body OZI-allowed strong decay properties for the high-lying $b\bar{b}$ resonances are given and discussed.
In Sec.~\ref{SDMix} and Sec.~\ref{aab}, the $S$-$D$ mixing and electron decay widths for
the vector resonances are evaluated, respectively.
In Sec.~\ref{crosss}, we further discuss the role of the high-lying $b\bar{b}$ vector resonances in the cross section of the $e^+e^-$ annihilation reaction. Finally, a summary is given in Sec.~\ref{SUM}.

\section{Unquenched quark model}\label{Fram}

\subsection{Hamiltonian for bare states}

The bare masses of $b\bar{b}$ states are calculated in a quenched semirelativistic potential model.
In this model, the effective Hamiltonian is
\begin{equation}\label{H0}
 H_0=\sqrt{\boldsymbol{q}_1^2+m_{b}^2}+\sqrt{\boldsymbol{q}_2^2+m_{\bar{b}}^2}+V_0(r)+V_{sd}(r),
\end{equation}
where $\boldsymbol{q}_1$ and $\boldsymbol{q}_2$ is the three momenta of quark and antiquark, respectively.
$r$ is the distance between quark and antiquarks, with $m_{b}$ and $m_{\bar{b}}$ representing their respective masses.
The $V_0(r)$ is spin-independent effective potential and adopts the well-known Cornell form~\cite{Eichten:1978tg}
\begin{eqnarray}
V_0(r)=-\frac{4}{3}\frac{\alpha_s}{r}+br+C_0,
\end{eqnarray}
which includes the color Coulomb interaction, linear confinement, and zero point energy $C_0$.
The spin-dependent part $V_{sd}(r)$ adopts the widely used form~\cite{Godfrey:1985xj, Barnes:2005pb}
\begin{eqnarray}\label{Vsd}
V_{sd}(r)&=&\frac{32\pi \alpha_s \cdot \sigma^3 e^{-\sigma^2 r^2}}{9 \sqrt{\pi^3}  m_{b}m_{\bar{b}}} \mathbf{S_q} \cdot \mathbf{S_{\bar{q}}}
\nonumber\\
 &&+\frac{4}{3}\frac{\alpha_s}{m_{b} m_{\bar{b}}}\frac{1}{r^3}\left(\frac{3\mathbf{S}_{q}\cdot \mathbf{r}\,\mathbf{S}_{{\bar{q}}}\cdot \mathbf{r}}{r^2}-\mathbf{S}_{q}\cdot\mathbf{S}_{{\bar{q}}}\right)+H_{LS}.
\end{eqnarray}
This part includes the spin-spin contact hyperfine potential, tensor potential, and spin-orbit interaction.
The spin-orbit interaction further decomposes into symmetric and antisymmetric terms, given as $H_{LS}=H_{sym}+H_{anti}$, with
\begin{eqnarray}\label{Lssym}
H_{sym}
=\frac{\mathbf{S_{+}\cdot L}}{2}\left[ \left(\frac{1}{2  m_b^2}+\frac{1}{2m_{\bar{b}}^2} \right) \left( \frac{4 \alpha_s}{3r^3}-\frac{b}{r}   \right)+\frac{8 \alpha_s}{3  m_b m_{\bar{b}} r^3} \right],
\end{eqnarray}
\begin{eqnarray}\label{Lsanti}
H_{anti} =\frac{\mathbf{S_{-}\cdot L}}{2}\left(\frac{1}{2  m_b^2}-\frac{1}{2m_{\bar{b}}^2} \right) \left( \frac{4 \alpha_s}{3r^3}-\frac{b}{r}   \right) .
\end{eqnarray}
In these equations, $\mathbf{L}$ is the orbital angular momentum of the $b\bar{b}$ system;
$\mathbf{S}_{\pm}\equiv \mathbf{S}_b\pm \mathbf{S}_{\bar{b}}$, $\mathbf{S}_b$ and $\mathbf{S}_{\bar{b}}$ are the spins of the quark and antiquark, respectively.
It should be emphasized that the antisymmetric term $H_{anti}$ has no contribution to the mass of $b\bar{b}$ states since the masses of quark and antiquark are equal.
The parameter set $\{\alpha_s,b,m_{b/\bar{b}},\sigma\}$ in the above potentials is determined by fitting the spectrum of low-lying bottomonium states.
In this work, we solve the radial Schr\"{o}dinger equation using the Gaussian expansion method~\cite{Hiyama:2003cu}.
The detailed discussions on the numerical calculation techniques involving the Gaussian expansion method can be found in our previous work~\cite{Ni:2021pce}.

\begin{figure}
	\centering \epsfxsize=4.5 cm \epsfbox{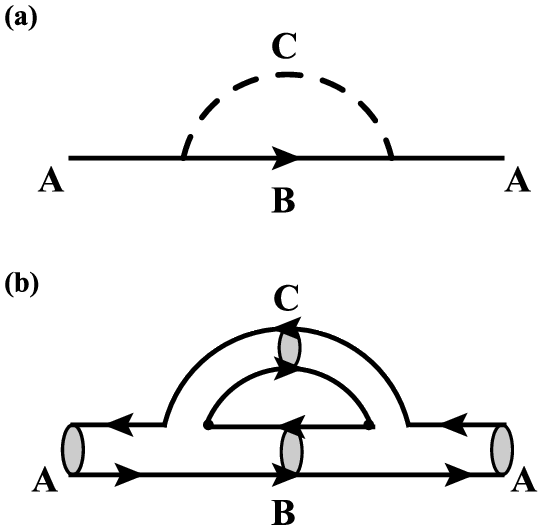} \vspace{-0.0 cm} \caption{The $BC$ hadronic loop coupled to a bare meson state $|A \rangle$, drawn at (a) the hadronic level and (b) the quark level.}\label{CCEFsFig}
\end{figure}

\subsection{Coupled-channel effects}\label{Coupled-channel effects}

A bare state $|A\rangle$ described within the quenched quark model can further couple to the two-hadron continuum $BC$ via hadronic loops, as illustrated in Fig.~\ref{CCEFsFig}.
The wave function of the physical state $| \Psi(M) \rangle$ with the mass eigenvalue $M$ is written as
\begin{equation}
| \Psi(M) \rangle =c_A(M) |A\rangle + \sum_{BC}\int c_{BC}(\boldsymbol{p},M)d^3\boldsymbol{p} |BC,\boldsymbol{p}\rangle ,
\end{equation}
where $\boldsymbol{p}=\boldsymbol{p}_B=-\boldsymbol{p}_C$ is final two-hadron relative momentum in the initial hadron static system, $c_A(M)$ and $c_{BC}(\boldsymbol{p}, M)$ denote the probability amplitudes of the bare valence state $|A\rangle$ and the continuum state $|BC,\boldsymbol{p}\rangle$, respectively.

The full Hamiltonian for the $| \Psi(M) \rangle$ is written as
\begin{equation}
H = H_0+H_c +H_I,
\end{equation}
where $H_0$ is the Hamiltonian for describing the bare state $|A\rangle$ based on the quark level, which has been given by Eq.(\ref{H0}).
$H_c$ is the Hamiltonian for describing the continuum state $|BC,\boldsymbol{p}\rangle$.
In principle, $H_c$ should include the kinetic energy of the continuum state and the interaction part between two continuum states.
However, due to the interactions between continuum states involving many unknown parameters, such as the interaction potentials of the $t$ and $u$ channels, the currently very limited $b\bar{b}$ bottomonium spectrum data is insufficient to extract information about these interactions.
Theoretically, we could incorporate these interactions into $H_I$; therefore, $H_c$ here only includes the kinetic energy term as follows,
\begin{equation}
H_{c}|BC,\boldsymbol{p}\rangle=E_{BC}|BC,\boldsymbol{p}\rangle,
\end{equation}
where $E_{BC}=\sqrt{m_B^2+p^2}+\sqrt{m_C^2+p^2}$ is the kinetic energy of the
$|BC,\boldsymbol{p}\rangle$ continuum state.

The term $H_I$ plays a crucial role in describing the coupling between the bare state $|A\rangle$ and the $|BC,\boldsymbol{p}\rangle$ continuum components.
In this work, we adopt the widely used $^3P_0$ model~\cite{Micu:1968mk, LeYaouanc:1972vsx, LeYaouanc:1973ldf} to depict the interaction $H_I$.
Within the $^3P_0$ model, the operator is expressed as
\begin{eqnarray}\label{T operator}
H_{I}&=&-3 \gamma \sqrt{96\pi } \sum_{m}\langle 1 m 1-m \mid 00\rangle \int d \boldsymbol{p}_{3} d \boldsymbol{p}_{4} \delta^{3}\left(\boldsymbol{p}_{3}+\boldsymbol{p}_{4}\right) \nonumber \\
&& \times \mathcal{Y}_{1}^{m}\left(\frac{\boldsymbol{p}_{3}-\boldsymbol{p}_{4}}{2}\right) \chi_{1-m}^{34} \phi_{0}^{34} \omega_{0}^{34} b_{3i}^{\dagger}\left(\boldsymbol{p}_{3}\right) d_{4j}^{\dagger}\left(\boldsymbol{p}_{4}\right),
\end{eqnarray}
where $\gamma$ is a dimensionless constant that denotes the strength of the quark-antiquark pair creation with momentum $\boldsymbol{p}_{3}$ and $\boldsymbol{p}_{4}$ from vacuum;
$b_{3i}^{\dagger}\left(\boldsymbol{p}_{3}\right)$ and $d_{4j}^{\dagger}\left(\boldsymbol{p}_{4}\right)$ are the creation operators for the quark and antiquark,
with indexes $i$ and $j$ for the SU(3)-color indices of the created quark and antiquark, respectively;
$\phi_{0}^{34}=\left (u\bar{u}+d\bar{d}+s\bar{s}\right ) /\sqrt{3}$ and $\omega_{0}^{34}= \delta_{ij}/\sqrt{3}$ correspond to flavor and color singlets, respectively;
$\chi_{1-m}^{34}$ is a spin triplet state;
$\mathcal{Y}_{\ell m}(\mathbf{k}) \equiv|\mathbf{k}|^{\ell} Y_{\ell m}\left(\theta_{\mathbf{k}}, \phi_{\mathbf{k}}\right)$ is the $\ell$-th solid harmonic polynomial.
For more details of the $^3P_0$ model can be found in our previous works~\cite{Gui:2018rvv,Li:2020xzs}.

The Schr\"{o}dinger-like equation of the composite hadron including coupled-channel effects expressed as
\begin{eqnarray}\label{coupled-channel equation}
&&\left( \begin{matrix}H_0~~~~~H_I
\\  H_I~~~~~H_c \end{matrix}\right)
\left( \begin{matrix} c_A(M) |A\rangle
\\ \sum_{BC}\int c_{BC}(\boldsymbol{p},M)d^3\boldsymbol{p} |BC,\boldsymbol{p}\rangle  \end{matrix}\right)~~~~~~~~~~~~\nonumber\\
&&~~~~~~~~~~~~~~~~~~~~~=
M \left( \begin{matrix} c_A(M) |A\rangle
\\  \sum_{BC}\int c_{BC}(\boldsymbol{p},M) d^3\boldsymbol{p}|BC,\boldsymbol{p}\rangle    \end{matrix}\right).
\end{eqnarray}
This equation gives rise to a system of coupled equations for $c_{A}$ and $c_{BC}(\boldsymbol{p}, M)$:
\begin{eqnarray}\label{coupled-channel equation1}
&&c_A(M) M_A+ \sum_{BC}\int c_{BC}(\boldsymbol{p},M)d^3\boldsymbol{p} \langle A | H_I | BC,\boldsymbol{p}\rangle \nonumber \\
&&~~~~~~~~~~~~~~~~~~~~~~~~~~~~~~~~~~~~~~~~~~~~~~
=
c_A(M) M ,
\end{eqnarray}
\begin{eqnarray}\label{coupled-channel equation2}
&& c_A(M)\langle BC,\boldsymbol{p'}| H_I | A \rangle +\int c_{BC}(\boldsymbol{p},M) d\boldsymbol{p'}^{3}E_{BC}\delta^{3}(\boldsymbol{p'}-\boldsymbol{p}) \nonumber \\
&&~~~~~~~~~~~~~~~~~~~~~~~~~~~~~~~~~~~~~~~~~~~~~~~~~~~~~
=
c_{BC}(\boldsymbol{p'},M) M .
\end{eqnarray}
Deriving $c_{BC}(\boldsymbol{p},M)$ from Eq.(\ref{coupled-channel equation2}), and substituting it into Eq.(\ref{coupled-channel equation1}), we obtain the hadronic self-energy as follows:
\begin{eqnarray}\label{gM}
g(M)&=&\sum_{BC} g_{BC}(M) \nonumber\\
&=&\sum_{BC} \int_0^{\infty}\frac{|f(\boldsymbol{p})|^2}{M-E_{BC}+i\epsilon} p^2 dp d\Omega_p \nonumber\\
&=&\sum_{B C} \int_{0}^{\infty} \frac{\overline{\left|\mathcal{M}_{A \rightarrow B C}(\boldsymbol{p})\right|^{2}}}{M-E_{B C}+i\epsilon} p^{2} d p,
\end{eqnarray}
where $f(\boldsymbol{p})=\langle BC,\boldsymbol{p}| H_I | A \rangle$, and $g_{BC}(M)$ represents the self-energy of the partial channel $BC$.

It should be emphasized that the infinitesimal term $i\epsilon~(\epsilon \to 0^+)$ is introduced in the self-energy function to properly deal with the kinematic region where $M > m_B + m_C$.
In this case, the self-energy function becomes complex:
\begin{equation}
g(M) = \text{Re}~g(M) - i\,\text{Im}~g(M).
\end{equation}
The real part, $\text{Re}~g(M)$, represents the mass shift induced by the coupling to intermediate hadronic channels.
It is given by
\begin{eqnarray}\label{ReDelta M}
\Delta M(M)
&=& \sum_{B C} \mathcal{P} \int_{0}^{\infty} \frac{\overline{\left|\mathcal{M}_{A \rightarrow B C}(\boldsymbol{p})\right|^{2}}}{M-E_{B C}} p^{2} d p,
\end{eqnarray}
where $\mathcal{P}$ is the principal part integral.
It should be emphasized that calculating the mass shift while considering an infinite number of channels is practically insurmountable.
In our calculations, we included all OZI-allowed two-body hadronic channels with mass thresholds below the bare $|A\rangle$ states.
In addition, we include channels whose thresholds lie slightly above the bare mass, within an upper limit of 100 MeV.
This choice is motivated by the fact that the bare masses predicted by potential models inherently carry uncertainties of several tens of MeV.
Due to these uncertainties, certain channels may correspond either to open-flavor thresholds or to nearby virtual channels.
Moreover, channels located close to the bare mass may couple to the bottomonium state via $S$-wave and induce sizable mass shifts.
Therefore, we include virtual meson pairs whose thresholds lie within 100 MeV above the predicted bare bottomonium mass to account for the most relevant coupled-channel effects.
This upper limit is consistent with the typical uncertainty scale of potential models and ensures the inclusion of channels most likely to induce significant mass shifts.
Unfortunately, with the current model and the limited experimental input, we can not distinguish the contributions of all the hadron channels, and in this situation, the best optimal choice is to include the several closest channels.
By contrast, virtual channels with thresholds far above the bare $|A\rangle$ state are treated separately using the once-subtracted dispersion relation formalism, following Refs.~\cite{Pennington:2007xr,Zhou:2011sp,Duan:2021alw,Ni:2023lvx,Deng:2023mza}.
These high-lying channels are numerous, and their effects vary slowly across different states.
As a result, their cumulative contribution can be well approximated by a constant, which can be absorbed into a redefinition of the bare mass.

Adopting the once-subtracted approach, the hadronic mass shift can be rewritten as~\cite{Ni:2023lvx,Deng:2023mza}
\begin{equation}\label{ReDelta MM0}
\Delta M(M)
= \sum_{BC} \mathcal{P} \int_{0}^{\infty}
\frac{\left(M_0-M\right)\overline{\left|\mathcal{M}_{A \rightarrow BC}(\boldsymbol{p})\right|^{2}}}{\left(M-E_{BC}\right)\left(M_0-E_{BC}\right)}
p^{2} d p.
\end{equation}
Here, $M_{0}$ represents the subtracted zero-point, which is typically the mass of a $b\bar{b}$ state nearly unaffected by intermediate hadronic loops.
In this work, we choose $\Upsilon(1S)$ as the subtracted zero-point with a value of $M_{0}=9460$ MeV.
It is important to emphasize that the choice of $M_0$ has minimal impact on our predictions of the mass shifts.
The reason for this lies on the factor $F\equiv(M_0-M)/[(M-E_{BC})(M_0-E_{BC})]$ depending on $M_{0}$ in the mass shift as shown in Eq.(\ref{ReDelta MM0}).
This factor can be expressed as $F=1/\left[(M-E_{BC})(1-\frac{M-E_{BC}}{M-M_0})\right]$.
Note that the mass shift $\Delta M(M)$ is primarily influenced by the integral around the pole position $M\sim E_{BC}$, where $M- E_{BC} \simeq 0$.
Since the subtracted zero-point $M_0$ is usually far below $E_{BC}$ (i.e., $M-M_0\sim E_{BC}-M_0\gg 0$), it follows that $\frac{M-E_{BC}}{M-M_0}\sim 0$.
Therefore, the factor $F$ is not significantly sensitive to the selection of $M_0$.
In other words, the mass shift $\Delta M(M)$ determined by the coupled-channel equation is not sensitive to the chosen subtraction point.

The coupled-channel equation is obtained from Eq.(\ref{coupled-channel equation1}), i.e.,
\begin{equation}\label{M=MA+Delta M}
M=M_A+\Delta M(M).
\end{equation}
Combining Eq.(\ref{ReDelta MM0}) and Eq.(\ref{M=MA+Delta M}), one can determine the physical mass $M$ and the mass shift $\Delta M(M)$ simultaneously.
However, it should be emphasized that when using the Eq.(\ref{ReDelta MM0}) to calculate the mass shift, the contribution in the higher $\boldsymbol{p}$ region may be nonphysical because the quark pair production rates via the non-perturbative interaction $H_I$ should be suppressed~\cite{Silvestre-Brac:1991qqx,Ni:2021pce,Zhong:2022cjx}.
The effective approach to overcome this difficulty is to introduce a suppressed factor into the transition amplitudes, i.e.,
\begin{eqnarray}\label{Form factor}
 \langle BC,\boldsymbol{p}| H_I | A \rangle
\to \langle BC,\boldsymbol{p}| H_I e^{-\boldsymbol{p}^2/(2\Lambda^2)} | A \rangle,
\end{eqnarray}
where $\Lambda$ is the cut-off parameter.
In this work, we adopt $\Lambda=0.78$ GeV to consist with our recent studies of heavy-light meson~\cite{Ni:2023lvx} and charmonium spectra~\cite{Deng:2023mza}.
This value is also comparable to $\Lambda=0.84$ GeV used in the literature~\cite{Ortega:2016mms,Ortega:2016pgg,Ortega:2020uvc}.
In fact, a similar suppression factor, $\sqrt{\Lambda^2/(\Lambda^2+\boldsymbol{q}^2)}$, is widely adopted in the study of hadron spectra and hadron-hadron interactions within the chiral quark model~\cite{Vijande:2004he, Brauer:1990kt, Zhang:1994pp, Yu:1995ag, Valcarce:2005em, Valcarce:2005rr}.
This suppression factor is related to the dynamics of quark-antiquark pair creation in the vacuum.
The cut-off parameter $\Lambda$ determines the scale at which chiral symmetry is broken.
The typical range of $\Lambda$ is around $0.8 \pm 0.2$ GeV~\cite{Vijande:2004he, Brauer:1990kt, Zhang:1994pp, Yu:1995ag, Valcarce:2005em, Valcarce:2005rr}.

If the physical mass of the initial hadron exceeds the threshold of the final states $B$ and $C$, i.e., $M > m_B + m_C$, the strong decay process $A \to BC$ will occur. The partial decay width for this process is given by
\begin{eqnarray}\label{width}
\Gamma_{BC}=2\pi \frac{|\boldsymbol{p}|E_BE_C}{M}\overline{\left|\mathcal{M}_{A \rightarrow B C}(\boldsymbol{p})\right|^{2}}.
\end{eqnarray}
According to the optical theorem, the imaginary part of the hadronic self-energy, $\text{Im}~g(M)$, can be related to the decay process through the cutting of the hadronic loop diagram, as shown in Fig.~\ref{CCEFsFig}.
This imaginary part is determined by the same transition amplitude $\mathcal{M}_{A \rightarrow BC}(\boldsymbol{p})$ that governs the decay and satisfies the relation $\text{Im}~g_{BC}(M) = \Gamma_{BC}/2$.
It is important to emphasize that both the mass shift and the decay width are determined by the same strong transition mechanism, i.e., the $^3P_0$ model, and by using consistent wave functions and parameters. Therefore, in our model, the real and imaginary parts of the self-energy are not independent, but instead emerge from a unified theoretical framework.

For the low-lying $b\bar{b}$ states, whose mass is smaller than the threshold of the lowest open-bottom channel, the derivative of the self-energy is associated with the probability of finding the bare state within the wave function of the physical state.
The component of the $b\bar{b}$ core is determined by~\cite{Weinberg:1962hj}
\begin{equation}\label{PA}
P_{b\bar{b}} \equiv |c_{A}|^2 =  \left(1- \frac{\partial }{\partial M} g (M)\bigg|_{M=M_{phy}}  \right)^{-1}.
\end{equation}
The above equation can also be readily derived from the normalization of the wave function of the bound state~\cite{Wang:2023snv}.
Furthermore, the probability of the presence of a $BC$ hadron pair is determined by
\begin{equation}\label{PBC}
P_{BC} \equiv |c_{BC}|^2 = -|c_{A}|^2\frac{\partial}{\partial M} g_{BC}(M) \bigg|_{M=M_{phy}} .
\end{equation}

On the other hand, for the high-lying $b\bar{b}$ resonances, who have one or more OZI-allowed two body strong decay channels, Eqs.~(\ref{PA}) and (\ref{PBC}) are invalid for calculating the components of physical states, because the wave function of the scattering state cannot be normalized directly.
To address this, a method known as the ``spectral density function'' provides an effective scheme for estimating the probability density of the bare state component.
This approach was originally introduced in Ref.~\cite{Bogdanova:1991zz} and further developed in Refs.\cite{Baru:2003qq, Kalashnikova:2005ui, Kalashnikova:2009gt, Baru:2010ww, Hanhart:2010wh, Giacosa:2007bn, Wang:2023snv}.
The spectral density near a resonance is given by
\begin{equation}\label{omegam1}
\omega_R(M)= 4\pi|\boldsymbol{p}|\frac{E_B E_C}{E_{BC}}|c_{A}(M)|^2.
\end{equation}
Here, $c_{A}(M)$ represents the probability density of a bare state within the spectral density function, determined as
\begin{equation}\label{cAM}
c_{A}(M)=\frac{f^{*}(\boldsymbol{p}_{M})}{M-M_{A}-g(M)}.
\end{equation}
Substituting Eq.(\ref{cAM}) into Eq.(\ref{omegam1}), one can obtain
\begin{equation}\label{omega2}
\omega_R(M)=\frac{1}{2\pi} \frac{\Gamma_{R}}{(M-M_{R})^2+\Gamma_{R}^2/4},
\end{equation}
where $M_{R}$ and $\Gamma_{R}\equiv\sum_{BC}\Gamma_{BC}$ denote the mass and width of the resonance, respectively.
It is easy to see that the spectral function in this case is the well-known Breit-Wigner form.
For a physical resonance state with a complex energy $M_{R} - i \Gamma_{R}/2$, the probability
of the $b\bar{b}$ core is estimated to be
\begin{equation}\label{omegadeltaM}
P_{b\bar{b}}=\int_{M_{R}-\Delta }^{M_{R}+\Delta }\omega(M)dM,
\end{equation}
where $\Delta$ is the integral interval what we choose.
If the spectral function around the resonance pole shows a typical Breit-Wigner line shape,
we take $\Delta=\Gamma_{R}$ as that usually adopted in the literature~\cite{Wang:2023snv}.
However, the line shape of the spectral function around the resonance pole is often distorted by the strong couplings of multiple channels to the bare $b\bar{b}$ state, which is not a typical Breit-Wigner form.
In this case, to obtain more reliable predictions, the integration interval is chosen by combining the line shape of the spectral functions near the physical poles.

\subsection{Model parameters}

In our calculations, the potential model Hamiltonian $H_0$ incorporates six parameters, $\{b,\alpha_{s}, m_{b}, \sigma, C_{0},r_c \}$, which define the interactions within the quenched framework.
Additionally, the effective interaction $H_I$ in the $^3P_0$ model includes a single parameter $\gamma$, representing the quark-antiquark pair creation strength in vacuum.

The parameters of the potential model are specified as follows:
$b = 0.180~\mathrm{GeV}^2$, $\alpha_s = 0.372$,  $m_b = 4.80$ GeV, $\sigma = 2.42$ GeV, and $C_0 = -35.0$ MeV.
In our model, $b$ is fixed at a typical value $0.180~\mathrm{GeV}^2$, commonly used in relativistic potential models~\cite{Godfrey:1985xj,Zeng:1994vj,Ebert:1997nk,Ebert:2009ua}.
The bottom quark mass $m_b$ is fixed at $4.80~\mathrm{GeV}$ to ensure consistency with our previous study on $B$ and $B_s$ mesons spectrum~\cite{Ni:2023lvx}.
Consequently, only three parameters, $\alpha_s$, $\sigma$, and $C_0$, are treated as adjustable.
These parameters are determined by minimizing a $\chi^2$ function defined with respect to the masses of well-established $b\bar{b}$ states far below the open-bottom threshold:
\begin{eqnarray}\label{chisquare}
\chi^2=\sum_{n} \frac{\left(M_{th}(n)-M_{exp}(n)\right)^2}{M_{err}(n)^2}.
\end{eqnarray}
Here, $M_{th}$, $M_{exp}$, and $M_{err}$ represent the theoretical mass, experimental mass, and experimental uncertainty, respectively.
During the fitting, $\alpha_s$ and $C_0$ mainly control the overall mass scale of low-lying bottomonium states.
The parameter $\sigma$, which appears in the spin-spin potential, is primarily constrained by the hyperfine splitting $M[\Upsilon(1S) - \eta_b(1S)]_{exp} \simeq 61~\mathrm{MeV}$.
The resulting comparison between theoretical predictions and experimental data is presented in Table~\ref{bbbar chisquare}, yielding a total $\chi^2$ value of 61.5.

It should be emphasized that stable solutions for some states cannot be obtained due to the singularity of the $1/r^3$ term in the spin-orbit and tensor potentials.
To address this issue, we introduce a cutoff distance $r_c$, following the method adopted in Refs.~\cite{Deng:2016ktl,Deng:2016stx,Li:2019tbn,Li:2019qsg,Li:2020xzs,Ni:2021pce}.
Specifically, in the region $r \leq r_c$, the divergent term $1/r^3$ is replaced by a constant $1/r_c^3$.
The $1^3P_0$ state is particularly sensitive to the choice of $r_c$, due to its large spin-orbit factor $\langle \mathbf{S}_+ \cdot \mathbf{L} \rangle$.
To determine the optimal value of $r_c$, we fit the observed fine splitting $M[\chi_{b1}(1P) - \chi_{b0}(1P)]_{exp} \simeq 33~\mathrm{MeV}$ using the perturbative approach as adopted in Refs.~\cite{Li:2019tbn,Li:2020xzs}.
This procedure yields $r_c = 0.101~\mathrm{fm}$.
In the perturbative approach, the spin-orbit and tensor interactions with the $1/r_c^3$ term are treated as small corrections.
The procedure involves four steps: (i) solving the system without $1/r_c^3$ terms to obtain the zeroth-order mass $M_0$ and the corresponding wave function; (ii) computing the mass correction $\delta M$ using the perturbative terms; (iii) adding the correction to obtain the total mass $M = M_0 + \delta M$; (iv) determining the cutoff parameter $r_c$ by fitting the total mass $M$.
While the perturbative method provides accurate mass estimates, it neglects the feedback of spin-dependent interactions on the wave function.
By introducing the cutoff $r_c$, these effects can be included nonperturbatively.
This allows both the mass and the wave function to be consistently corrected by the spin-orbit and tensor potentials.

In our calculations of strong decay widths, the wave functions of the initial $b\bar{b}$ states are adopted from our quark model predictions.
For the final $B$ and $B_s$ mesons in a decay process, the wave functions are adopted from the predictions of our previously work~\cite{Ni:2023lvx}.
To be consistent with the mass calculations within the potential model, the bottom quark mass is adopted as $m_{b}=4.80$ GeV.
For the light quark masses, we adopt $m_{u,d}=0.35$ GeV and $m_{s}=0.55$ GeV, which are consistent with those adopted for the calculations of the $B$- and $B_s$-meson spectra~\cite{Ni:2023lvx}.
For the well-established hadrons involving in the final states, the masses are adopted the averaged values from the PDG~\cite{ParticleDataGroup:2024cfk}, while for the missing states, the masses are adopted the theoretical predictions in a recent work~\cite{Ni:2023lvx}.
The creation quark pair strength $\gamma=0.482$ within the $^3P_0$ model is determined by fitting the measured width of $\Upsilon(10580) \to \bar{B}B=20.5$ MeV~\cite{ParticleDataGroup:2024cfk}.

\begin{table}
	\begin{center}
		\caption{The model parameter determination by fitting the masses of the well-established states with
the minimum $\chi^2$ method. The errors of the observed masses are uniformly taken as $M_{err}=5$ MeV.
The unit of mass is MeV. }
		\label{bbbar chisquare}
		\renewcommand\arraystretch{1.15}
		\tabcolsep=0.180cm
		\scalebox{1.0}{
			\begin{tabular}{cccccccccccccccccccc}
				\bottomrule[1.0pt]\bottomrule[1.0pt]
				Number  &$n^{2S+1}L_J$    &Observed State   &$M_{exp}$~\cite{ParticleDataGroup:2024cfk}    &$M_{err}$  &$M_{th}$  \\
				\bottomrule[0.5pt]	
				1&$1^1S_0$    &$\eta_{b}(1S)$   &$9399$    &$5$  &$9399$  \\
				2&$1^3S_1$    &$\Upsilon(1S)$   &$9460$    &$5$  &$9460$  \\
				3&$2^1S_0$    &$\eta_{b}(2S)$   &$9999$    &$5$  &$9991$  \\
				4&$2^3S_1$    &$\Upsilon(2S)$   &$10023$   &$5$  &$10013$  \\
				5&$3^3S_1$    &$\Upsilon(3S)$   &$10355$   &$5$  &$10348$  \\
				6&$1^3P_0$    &$\chi_{b0}(1P)$  &$9859$    &$5$  &$9877$  \\
				7&$1^1P_1$    &$h_{b}(1P)$      &$9899$    &$5$  &$9916$  \\
				8&$1^3P_1$    &$\chi_{b1}(1P)$  &$9893$    &$5$  &$9910$  \\
				9&$1^3P_2$    &$\chi_{b2}(1P)$  &$9912$    &$5$  &$9930$  \\
				10&$2^3P_0$    &$\chi_{b0}(2P)$  &$10233$   &$5$  &$10233$  \\
				11&$2^1P_1$    &$h_{b}(2P)$      &$10260$   &$5$  &$10261$  \\
				12&$2^3P_1$    &$\chi_{b1}(2P)$  &$10255$   &$5$  &$10258$  \\
				13&$2^3P_2$    &$\chi_{b2}(2P)$  &$10269$   &$5$  &$10273$  \\
				14&$1^3D_2$    &$\Upsilon_2(1D)$ &$10164$   &$5$  &$10162$  \\
				\bottomrule[0.5pt]
				$\chi^2$ & &                 &          &     &$61.5$  \\
				\bottomrule[1.0pt]\bottomrule[1.0pt]
		\end{tabular}}
	\end{center}
\end{table}

\begin{table*}
	\begin{center}
		\caption{Our theoretical masses (MeV) are compared with experimental data and predictions from other quark models. The mass spectrum is presented within both quenched $(Q)$ and unquenched $(UQ)$ frameworks. The mass shifts due to the coupled-channel effects are denoted as $\Delta M$. For the low-lying states having no strong decay two-body open-bottom meson channels, the mass corrections
due to the coupled-channel effects are absorbed in the mass parameters with the once-subtracted method, thus no mass shifts are given.
In the unquenched pictures, the $b\bar{b}$-core components evaluated for the resonances are denoted as $P_{b\bar{b}}~(\%)$.}
		\label{MassSpectrumTable}
		\renewcommand\arraystretch{1.15}
		\tabcolsep=0.150cm
		\scalebox{1.0}{
			\begin{tabular}{ccccccccccccccccccccccc}
				\bottomrule[1.0pt]\bottomrule[1.0pt]
				$n^{2S+1}L_{J}$
				&   $J^{PC}$
				&    $Q$
				&    $\Delta M$
				&    $UQ$
				&    $P_{b\bar{b}}$
				&    Exp.~\cite{ParticleDataGroup:2024cfk}
				&    LC \cite{Li:2009nr}
				&   EFG \cite{Ebert:2011jc}
				&   WSLM \cite{Wang:2018rjg}
				&   DLGZ \cite{Deng:2016ktl}
				&   GI \cite{Godfrey:1985xj}\\
				\bottomrule[0.5pt]	
			          $ 3^{1}S_{0}$      &     $0^{-+}$ &     $10334$     &  $- $       &  $10334$       &$-$        &     $...$       &$10330$ &     $10329$ &   $10336$ &     $10326$ &     $10336$ \tabularnewline	
				$ 4^{1}S_{0}$      &     $0^{-+}$ &     $10608$     &  $-37$      &  $10571$       &$68$       &     $...$       &$10595$ &     $10573$ &   $10597$ &     $10584$ &     $10623$ \tabularnewline		
				$ 4^{3}S_{1}$      &     $1^{--}$ &     $10619$     &  $-40$      &  $10579$       &$61$       &     $10579$     &$10611$ &     $10586$ &   $10612$ &     $10597$ &     $10635$ \tabularnewline			
				$ 5^{1}S_{0}$      &     $0^{-+}$ &     $10846$     &  $-6$       &  $10840$       &$53$       &     $...$       &$10817$ &     $10851$ &   $10810$ &     $10800$ &     $10869$ \tabularnewline
				$ 5^{3}S_{1}$      &     $1^{--}$ &     $10856$     &  $-2$       &  $10854$       &$75$       &     $10885$     &$10831$ &     $10869$ &   $10822$ &     $10811$ &     $10878$ \tabularnewline
				$ 6^{1}S_{0}$      &     $0^{-+}$ &     $11062$     &  $-55$      &  $11007$       &$35$       &     $...$       &$11011$ &     $11061$ &   $10991$ &     $10988$ &     $11097$ \tabularnewline
				$ 6^{3}S_{1}$      &     $1^{--}$ &     $11071$     &  $-56$      &  $11015$       &$37$       &     $11000$     &$11023$ &     $11088$ &   $11001$ &     $10997$ &     $11102$ \tabularnewline
				$ 7^{1}S_{0}$      &     $0^{-+}$ &     $11262$     &  $-8$       &  $11254$       &$60$       &     $...$       &$11183$ &     $...$   &   $11149$ &     $...$   &     $...$ \tabularnewline
				$ 7^{3}S_{1}$      &     $1^{--}$ &     $11269$     &  $-10$      &  $11259$       &$55$       &     $...$       &$11193$ &     $...$   &   $11157$ &     $...$   &     $...$ \tabularnewline
				\bottomrule[0.3pt]
				$ 3^{3}P_{0}$      &     $0^{++}$ &     $10514$     & $-16$       &  $10498$       &$84$       &     $...$       &$10502$ &     $10521$ &   $10513$ &     $10490$ &     $10522$ \tabularnewline			
				$ 3^{1}P_{1}$      &     $1^{+-}$ &     $10538$     & $-17$       &  $10521$       &$88$       &     $...$       &$10529$ &     $10544$ &   $10530$ &     $10519$ &     $10541$ \tabularnewline
				$ 3^{3}P_{1}$      &     $1^{++}$ &     $10536$     & $-22$       &  $10514$       &$86$       &     $10513$     &$10524$ &     $10541$ &   $10527$ &     $10515$ &     $10538$ \tabularnewline
				$ 3^{3}P_{2}$      &     $2^{++}$ &     $10550$     &  $-27$      &  $10523$       &$85$       &     $10524$     &$10540$ &     $10550$ &   $10539$ &     $10528$ &     $10550$ \tabularnewline							          				
				$ 4^{3}P_{0}$      &     $0^{++}$ &     $10758$     &  $+14$      &  $10772$       &$53$       &     $...$       &$10732$ &     $10781$ &   $10736$ &     $...$   &     $10775$ \tabularnewline			
				$ 4^{1}P_{1}$      &     $1^{+-}$ &     $10780$     &  $-1$       &  $10779$       &$50$       &     $...$       &$10757$ &     $10804$ &   $10751$ &     $...$   &     $10790$ \tabularnewline
				$ 4^{3}P_{1}$      &     $1^{++}$ &     $10777$     &  $-2$       &  $10775$       &$49$       &     $...$       &$10753$ &     $10802$ &   $10749$ &     $...$   &     $10788$ \tabularnewline
				$ 4^{3}P_{2}$      &     $2^{++}$ &     $10790$     &  $+1$       &  $10791$       &$60$       &     $...$       &$10767$ &     $10812$ &   $10758$ &     $...$   &     $10798$ \tabularnewline							          								          				 
				$ 5^{3}P_{0}$      &     $0^{++}$ &     $10978$     &  $-21$      &  $10957$       &$74$       &     $...$       &$10933$ &     $...$   &   $10926$ &     $...$   &     $11004$ \tabularnewline			
				$ 5^{1}P_{1}$      &     $1^{+-}$ &     $10998$     &  $-40$      &  $10958$       &$51$       &     $...$       &$10955$ &     $...$   &   $10938$ &     $...$   &     $11016$ \tabularnewline
				$ 5^{3}P_{1}$      &     $1^{++}$ &     $10996$     &  $-30$      &  $10966$       &$58$       &     $...$       &$10951$ &     $...$   &   $10936$ &     $...$   &     $11014$ \tabularnewline
				$ 5^{3}P_{2}$      &     $2^{++}$ &     $11008$     &  $-32$      &  $10976$       &$47$       &     $...$       &$10965$ &     $...$   &   $10944$ &     $...$   &     $11022$ \tabularnewline							          								          				 
				$ 6^{3}P_{0}$      &     $0^{++}$ &     $11182$     &  $+25$      &  $11207$       &$46$       &     $...$       &$...$   &     $...$   &   $11090$ &     $...$   &     $...$ \tabularnewline			
				$ 6^{1}P_{1}$      &     $1^{+-}$ &     $11199$     &  $+6$       &  $11205$       &$44$       &     $...$       &$...$   &     $...$   &   $11101$ &     $...$   &     $...$ \tabularnewline
				$ 6^{3}P_{1}$      &     $1^{++}$ &     $11198$     &  $+5$       &  $11203$       &$56$       &     $...$       &$...$   &     $...$   &   $11099$ &     $...$   &     $...$ \tabularnewline
				$ 6^{3}P_{2}$      &     $2^{++}$ &     $11209$     &  $+7$       &  $11216$       &$37$       &     $...$       &$...$   &     $...$   &   $11106$ &     $...$   &     $...$ \tabularnewline							          								          				 
				\bottomrule[0.3pt]
				$ 1^{3}D_{1}$      &     $1^{--}$ &     $10154$     &  $- $        &  $10154$       &$-$          &     $...$       &$10145$ &     $10154$ &   $10153$ &     $10146$ &     $10138$ \tabularnewline
				$ 1^{1}D_{2}$      &     $2^{-+}$ &     $10163$     &  $- $        &  $10163$       &$-$          &     $...$       &$10152$ &     $10163$ &   $10163$ &     $10153$ &     $10148$ \tabularnewline	
				$ 1^{3}D_{3}$      &     $3^{--}$ &     $10167$     &  $- $        &  $10167$       &$-$          &     $...$       &$10156$ &     $10166$ &   $10170$ &     $10157$ &     $10155$ \tabularnewline						 				
				$ 2^{3}D_{1}$      &     $1^{--}$ &     $10440$     &  $- $        &  $10440$       &$-$          &     $...$       &$10432$ &     $10435$ &   $10442$ &     $10425$ &     $10441$ \tabularnewline       				
				$ 2^{1}D_{2}$      &     $2^{-+}$ &     $10449$     &  $- $        &  $10449$       &$-$          &     $...$       &$10439$ &     $10445$ &   $10450$ &     $10432$ &     $10450$ \tabularnewline       				
				$ 2^{3}D_{2}$      &     $2^{--}$ &     $10448$     &  $- $        &  $10448$       &$-$          &     $...$       &$10438$ &     $10443$ &   $10450$ &     $10432$ &     $10449$ \tabularnewline       				
				$ 2^{3}D_{3}$      &     $3^{--}$ &     $10454$     &  $- $        &  $10454$       &$-$          &     $...$       &$10442$ &     $10449$ &   $10456$ &     $10436$ &     $10455$ \tabularnewline       				
				$ 3^{3}D_{1}$      &     $1^{--}$ &     $10688$     &  $-18$      &  $10670$       &$62$       &     $...$       &$10670$ &     $10704$ &   $10675$ &     $...$   &     $10698$ \tabularnewline				
				$ 3^{1}D_{2}$      &     $2^{-+}$ &     $10697$     &  $-6$       &  $10691$       &$91$       &     $...$       &$10677$ &     $10713$ &   $10681$ &     $...$   &     $10706$ \tabularnewline  	 			
				$ 3^{3}D_{2}$      &     $2^{--}$ &     $10696$     &  $-11$      &  $10685$       &$76$       &     $...$       &$10676$ &     $10711$ &   $10681$ &     $...$   &     $10705$ \tabularnewline
				$ 3^{3}D_{3}$      &     $3^{--}$ &     $10702$     &  $-52/+16$  &  $10650/10718$ &$30/60$    &     $...$       &$10680$ &     $10717$ &   $10686$ &     $...$   &     $10711$ \tabularnewline
				$ 4^{3}D_{1}$      &     $1^{--}$ &     $10911$     &  $-4$       &  $10907$       &$68$       &     $...$       &$10877$ &     $10949$ &   $10871$ &     $...$   &     $10928$ \tabularnewline			
				$ 4^{1}D_{2}$      &     $2^{-+}$ &     $10920$     &  $+6$       &  $10926$       &$70$       &     $...$       &$10883$ &     $10959$ &   $10876$ &     $...$   &     $10935$\tabularnewline				
				$ 4^{3}D_{2}$      &     $2^{--}$ &     $10919$     &  $-6$       &  $10913$       &$72$       &     $...$       &$10882$ &     $10957$ &   $10876$ &     $...$   &     $10934$ \tabularnewline
				$ 4^{3}D_{3}$      &     $3^{--}$ &     $10925$     &  $-7$       &  $10918$       &$71$       &     $...$       &$10886$ &     $10963$ &   $10880$ &     $...$   &     $10939$\tabularnewline				
				$ 5^{3}D_{1}$      &     $1^{--}$ &     $11116$     &  $-18$      &  $11098$       &$33$       &     $...$       &$11060$ &     $...$   &   $11041$ &     $...$   &     $...$\tabularnewline				
				$ 5^{1}D_{2}$      &     $2^{-+}$ &     $11126$     &  $-62/-38$  &  $11064/11088$ &$10/33$    &     $...$       &$11066$ &     $...$   &   $11046$ &     $...$   &     $...$\tabularnewline				
				$ 5^{3}D_{2}$      &     $2^{--}$ &     $11125$     &  $-44$      &  $11081$       &$34$       &     $...$       &$11065$ &     $...$   &   $11045$ &     $...$   &     $...$\tabularnewline
				$ 5^{3}D_{3}$      &     $3^{--}$ &     $11131$     &  $-67/-41$  &  $11064/11090$ &$12/21$    &     $...$       &$11069$ &     $...$   &   $11049$ &     $...$   &     $...$\tabularnewline
				$ 6^{3}D_{1}$      &     $1^{--}$ &     $11308$     &  $-6$       &  $11302$       &$70$       &     $...$       &$...$ &       $...$   &   $...$   &     $...$   &     $...$\tabularnewline				
				$ 6^{1}D_{2}$      &     $2^{-+}$ &     $11318$     &  $-7$       &  $11311$       &$76$       &     $...$       &$...$ &       $...$   &   $...$   &     $...$   &     $...$\tabularnewline				
				$ 6^{3}D_{2}$      &     $2^{--}$ &     $11317$     &  $-5$       &  $11312$       &$72$       &     $...$       &$...$ &       $...$   &   $...$   &     $...$   &     $...$\tabularnewline
				$ 6^{3}D_{3}$      &     $3^{--}$ &     $11323$     &  $-11$      &  $11312$       &$64$       &     $...$       &$...$ &       $...$   &   $...$   &     $...$   &     $...$\tabularnewline
				\bottomrule[1.0pt]\bottomrule[1.0pt]
		\end{tabular}}
	\end{center}
\end{table*}

\begin{figure*}
	\centering \epsfxsize=12.4 cm \epsfbox{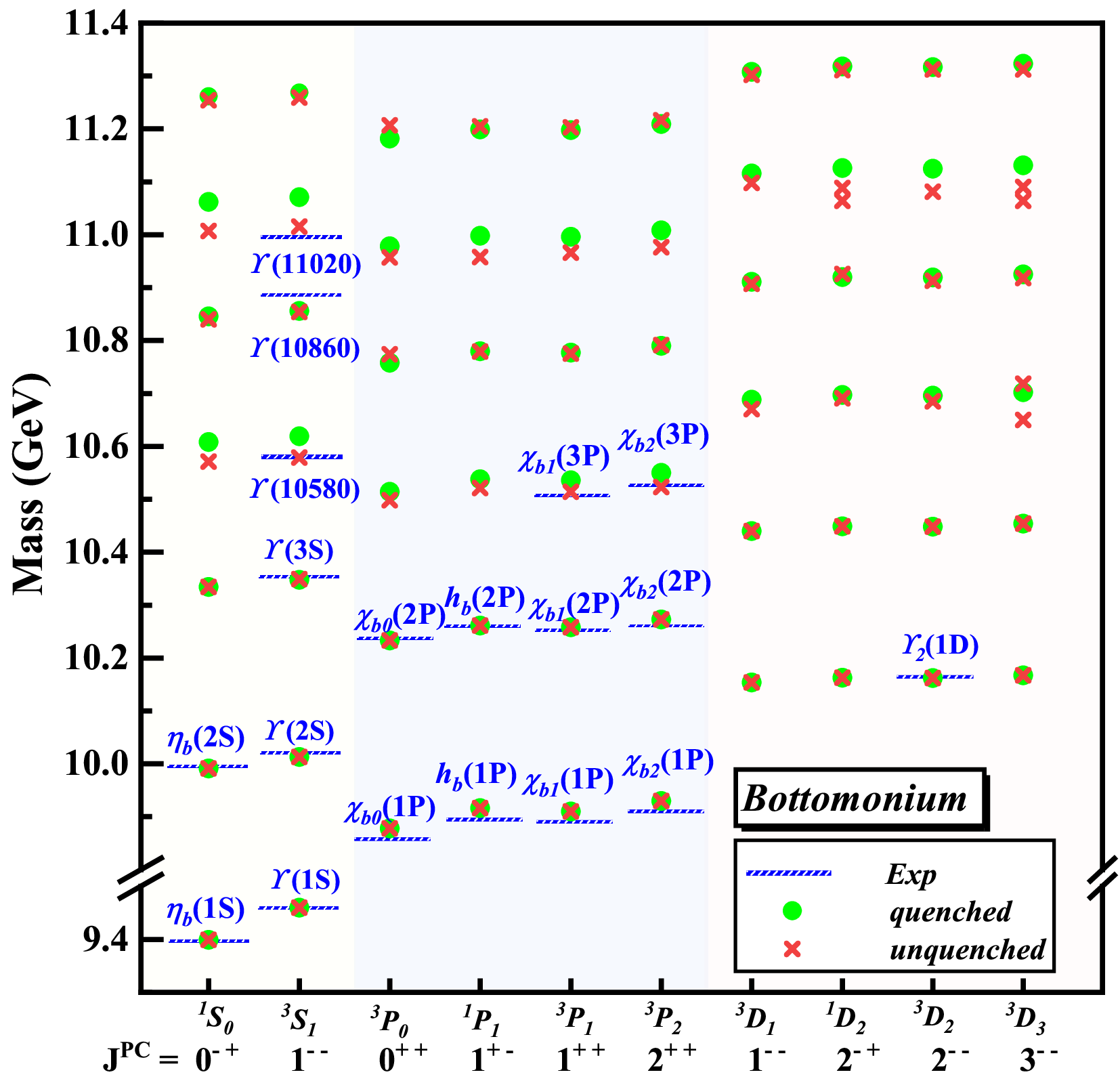} \vspace{-0.0 cm} \caption{Bottomonium spectrum compared with the observations. The well established states are labeled with blue color. The results obtained within the unquenched and quenched quark models are labeled by red crosses and solid green circles, respectively.
	}\label{MassSpectrumFig}
\end{figure*}

\begin{table*}
\begin{center}
\caption{Our predicted hyperfine and fine splittings (MeV) in both quenched ($Q$) and unquenched ($UQ$) frameworks compared
with the data and other model predictions. The data are taken from the RPP~\cite{ParticleDataGroup:2024cfk}.}\label{Masssplit}
\tabcolsep=0.30cm
\begin{tabular}{ccccccccc}
\hline\hline
 $\Delta m$
 &$Q$
 &$UQ$
 &Exp.~\cite{ParticleDataGroup:2024cfk}
 &GM~\cite{Godfrey:2015dia}
 &EFG~\cite{Ebert:2011jc}
 &LC~\cite{Li:2009nr}
 &DLGZ~\cite{Deng:2016ktl}
 &WSLM~\cite{Wang:2018rjg} \\
 \hline
$m(1^3S_{1})$-$m(1^1S_{0})$   &$61$      &$61$     &$62.3\pm 3.2$            &$63$     &$62$    &$71$      &$70$    &$65$      \\
$m(2^3S_{1})$-$m(2^1S_{0})$   &$22$      &$22$     &$24\pm 4$                &$27$     &$33$    &$29$      &$25$    &$28$      \\
$m(3^3S_{1})$-$m(3^1S_{0})$   &$14$      &$14$     &$$                       &$18$     &$26$    &$21$      &$17$    &$20$      \\
$m(4^3S_{1})$-$m(4^1S_{0})$   &$11$      &$8$      &$$                       &$12$     &$13$    &$16$      &$13$    &$15$      \\
$m(5^3S_{1})$-$m(5^1S_{0})$   &$10$      &$14$     &$$                       &$9$      &$18$    &$14$      &$11$    &$12$      \\
$m(6^3S_{1})$-$m(6^1S_{0})$   &$9$       &$8$      &$$                       &$5$      &$27$    &$12$      &$9$     &$10$     \\
$m(7^3S_{1})$-$m(7^1S_{0})$   &$7$       &$5$      &$$                       &$...$    &$...$   &$10$      &$...$   &$8$      \\
$m(3^3S_{1})$-$m(2^3S_{1})$   &$335$     &$335$    &$331.5\pm 0.13$          &$351$    &$332$   &$335$     &$328$   &$339$         \\
$m(4^3S_{1})$-$m(3^3S_{1})$   &$271$     &$231$    &$224.3\pm 1.7$           &$280$    &$231$   &$260$     &$254$   &$256$         \\
$m(5^3S_{1})$-$m(4^3S_{1})$   &$237$     &$275$    &$305.8^{+3.8}_{-2.8} $   &$243$    &$283$   &$220$     &$214$   &$210$        \\
$m(6^3S_{1})$-$m(5^3S_{1})$   &$215$     &$161$    &$114.8^{+6.6}_{-5.6} $   &$224$    &$219$   &$192$     &$186$   &$179$        \\
$m(7^3S_{1})$-$m(6^3S_{1})$   &$198$     &$244$    &                         &$...$    &$...$   &$170$     &$...$   &$156$        \\
$m(1^3P_{2})$-$m(1^3P_{1})$   &$20$      &$20$     &$19.10\pm 0.25$          &$21$     &$20$    &$21$      &$18$    &$21$     \\
$m(1^3P_{1})$-$m(1^3P_{0})$   &$33$      &$33$     &$32.5\pm 0.9$            &$29$     &$33$    &$32$      &$39$    &$31$     \\
$m(2^3P_{2})$-$m(2^3P_{1})$   &$15$      &$15$     &$13.10\pm 0.24$          &$15$     &$13$    &$18$      &$15$    &$14$     \\
$m(2^3P_{1})$-$m(2^3P_{0})$   &$25$      &$25$     &$23.5\pm 1.0$            &$20$     &$22$    &$25$      &$29$    &$20$     \\
$m(3^3P_{2})$-$m(3^3P_{1})$   &$14$      &$9$      &$10.6\pm 1.5$            &$12$     &$9$     &$16$      &$13$    &$12$         \\
$m(3^3P_{1})$-$m(3^3P_{0})$   &$22$      &$16$     &                         &$16$     &$20$    &$22$      &$25$    &$14$          \\
\hline\hline
\end{tabular}
\end{center}
\end{table*}


\section{Mass spectrum}\label{MDiscussion}

The mass spectrum obtained from the unquenched quark model is given Table~\ref{MassSpectrumTable}, where some other model predictions are also listed for comparison.
Furthermore, the theoretical mass spectrum compared with the data is plotted in Fig.~\ref{MassSpectrumFig}.
It is shown that the observed bottomonium spectrum below the $\bar{B}B$ threshold can be well described within the quark model.
Above the $\bar{B}B$ threshold, four vector resonances $\Upsilon(10580)$, $\Upsilon(10753)$,
$\Upsilon(10860)$, and $\Upsilon(11020)$ observed in experiments.
The masses of the $\Upsilon(10580)$ and $\Upsilon(11020)$ can be well understood in theory
with the $\Upsilon(4S)$ and $\Upsilon(6S)$ assignments in the unquenched quark model framework.
Assigning the $\Upsilon(10860)$ to the $\Upsilon(5S)$ state, our predicted mass
is about $30$ MeV lower than the PDG average value $M_{exp}=10885$ MeV.
At last, the $\Upsilon(10753)$ cannot be reasonably explained with any vector bottomonium states.
If assigning it to the $3^3D_1$ state, our predicted mass is about 100 MeV lower than the measured value.

Furthermore, in Table~\ref{Masssplit}, we present our predictions for the hyperfine splittings of some $S$-wave states and the fine splittings of some $P$-wave states.
For comparison, we also list the data and the predictions from other works~\cite{Godfrey:2015dia, Ebert:2011jc, Li:2009nr, Deng:2016ktl, Wang:2018rjg} in the same table.
It is seen that our predicted mass splittings are in good agreement with the observations.
It is worth noting that, after including the unquenched coupled-channel effects, our predicted splittings of $m(4^3S_1)-m(3^3S_1)$, $m(5^3S_1)-m(4^3S_1)$, and $m(6^3S_1)-m(5^3S_1)$ are more agreement with the observations.

\subsection{Mass shift due to coupled channel effects}

The bare masses of the high $b\bar{b}$ states may be significantly shifted
to the physical points due to some strong couplings to their nearby strong decay channels.
In Table~\ref{MassSpectrumTable}, we give our predictions of
the mass shifts contributed by these coupled-channel effects.
It is found that these effects may contribute a sizeable
mass correction ($\sim$ tens MeV) to some high-lying states.
It should be highlight for the understanding the $\Upsilon(10580)$ and $\Upsilon(11020)$ states.
It is based on mass shifts due to coupled-channel channel effects, the masses of $\Upsilon(4S)$ and $\Upsilon(6S)$ experience shifts of about $-40$ and $-60$ MeV, respectively.
In contrast to the original quark model predictions that did not account for the coupled-channel effects~\cite{Ebert:2011jc,Godfrey:1985xj}, these mass shifts align much more closely with experimental measurements, providing a solid theoretical foundation for understanding these two states.
Latter, from their components analysis, $\Upsilon(11020)$ indeed shows rich continuum components.
The contributions of each coupled channels to the mass shifts of the bare states are given in Tables~\ref{4S5S6S}-\ref{6D}.

\subsection{Components of physical states}

The physical states with high masses may contain significant continuum components of open-bottom meson pairs.
Within the unquenched framework, we estimate the $b\bar{b}$ core components for the $nS$ $(n=4,5,6,7)$, $nP$ and $nD$ $(n=3,4,5,6)$ states.
The results are given in Table~\ref{MassSpectrumTable}.
It should be pointed out that for the $\eta_b(4S)$ and $3P$ states, since the masses are slightly below the lowest threshold of their opened-bottom decay channels, their $b\bar{b}$ core and continuum components are estimated with Eqs.~(\ref{PA}) and (\ref{PBC}).
While for the other high mass resonances which have OZI allowed two-body strong decay channels, their $b\bar{b}$ core components are evaluated with Eq.~(\ref{omegadeltaM}) by combining the spectral density functions.

Our results show that for the well established resonance $\Upsilon(10580)$, besides the dominant $b\bar{b}$ component, $P_{b\bar{b}} \simeq 60~\%$, the $\bar{B}B^*+\bar{B}^*B^*$ component may reach up to a fairly large value $\sim40~\%$.
While the $\Upsilon(11020)$ resonance may be a continuum rich state,
the $b\bar{b}$-core component is only $\sim40~\%$.
Generally, from Table~\ref{MassSpectrumTable}, it is found that
the $4S$, $5S$ and $7S$ states may be $b\bar{b}$ dominant states with a component of
$P_{b\bar{b}} \simeq 55-75~\%$, however, for the $6S$ states the $b\bar{b}$ component is estimated to be only a small value $\sim40~\%$.
The small $b\bar{b}$ core components are also predicted
for the $6P$ and $5D$ states, which are in the range of $10-40~\%$.

\subsection{Two pole structures}

It is interesting to note that two pole structures arise from the $\Upsilon_3(3D)$, $\Upsilon_3(5D)$, and $\eta_{b2}(5D)$ states when coupled-channel effects are considered.
This phenomenon is similar to that observed in the case of $\chi_{c1}(2P)$~\cite{Deng:2023mza}.


Firstly, let's focus on the $\Upsilon_3(3D)$ state.
Nearby the threshold of $\bar{B}^*B^*$ channel, since the $\Upsilon_3(3D)$ couples strongly to this channel, as shown in the upper panel of Fig.~\ref{bbbarOthersDwaveMSandSF}, there exists a notable dip the mass shift function $\Delta M(M)$.
By solving the coupled-channel Eq.~(\ref{M=MA+Delta M}), we obtain two physical solutions with masses $10650$ and $10718$ MeV,
which two solutions correspond to the narrow and broad resonance structures in the spectral function as shown in the lower panel of Fig.~\ref{bbbarOthersDwaveMSandSF}, respectively.
This indicates that there may exist two physical states $\Upsilon_3(10650)$ and $\Upsilon_3(10718)$ arising from the bare state $\Upsilon_3(3D)$.
The narrow resonance $\Upsilon_3(10650)$ lies slightly below the $\bar{B}^*B^*$ threshold.
This state illustrated as a $\bar{B}^*B^*$ dominant state, since the $b\bar{b}$-core component is only $P_{b\bar{b}} \simeq 30~\%$.
On the other hand, the broad resonance $\Upsilon_3(10718)$ is a $b\bar{b}$ dominant state, the $b\bar{b}$-core component is estimated to be $P_{b\bar{b}} \simeq 60~\%$.
It should be noted that an unphysical solution can be found just above the $\bar{B}^*B^*$ threshold, as shown in the upper panel of Fig.~\ref{bbbarOthersDwaveMSandSF}.
This solution does not correspond to any resonance structure in the spectral function, as evidenced by the lower panel of Fig.~\ref{bbbarOthersDwaveMSandSF}.

Then we focus on the $\Upsilon_3(5D)$ and $\eta_{b2}(5D)$ states.
Both $\Upsilon_3(5D)$ and $\eta_{b2}(5D)$ couple strongly to the $\bar{B}^*B^*_2(5747)$ channel, the strong coupled-channel effects from this channel will cause a notable dip in the mass shift function of each bare state nearby their corresponding mass thresholds.
By solving Eq.~(\ref{M=MA+Delta M}) and combining the spectral function as shown in the third panel of Fig.~\ref{bbbarOthersDwaveMSandSF}, we obtain two physical states $\Upsilon_3(11064)$ and $\Upsilon_3(11090)$ arising from the $\Upsilon_3(5D)$, and other two physical states $\eta_{b2}(11064)$ and $\eta_{b2}(11088)$ arising from the $\eta_{b2}(5D)$.
Furthermore, it can be seen that the $\Upsilon_3(11064)$ and $\eta_{b2}(11064)$ are narrow resonances, while the $\Upsilon_3(11090)$ and $\eta_{b2}(11088)$ are broad resonances.
This is due to the opening of decay channels for the $S$-wave mode, which significantly contributes to the decay widths.
Both the $\Upsilon_3(11064)$ and $\eta_{b2}(11064)$ may be $\bar{B}^*B^*(5747)$ dominant states, their $b\bar{b}$-core components are estimated
to be $P_{b\bar{b}} \simeq 12~\%$ and $10~\%$, respectively.
For the two broad states $\Upsilon_3(11090)$ and $\eta_{b2}(11088)$, the $b\bar{b}$-core components are estimated to be $P_{b\bar{b}} \simeq 21~\%$ and $33~\%$, respectively, their components are dominated by the continuum states of various open-bottom meson pairs.

\section{Strong decays}\label{DDiscussion}

For the high-lying states with a mass above the $B\bar{B}$ threshold, the OZI-allowed two body strong decay channels are open.
The partial and total strong decay widths for these high-lying $nS$ $(n=5,6,7)$, $nP$ $(n=4,5,6)$, and $nD$$(n=3,4,5,6)$ states are estimated by using Eq.~(\ref{width}).
Our results have been given in Tables~\ref{4S5S6S}-\ref{6D}.
It should be mentioned that in this work we take the measured width $\Gamma_{exp}\simeq 20.5$ MeV~\cite{ParticleDataGroup:2024cfk} of the $\Upsilon(10580)$ as an input to determine the strength of quark pairs creation in the $^3P_0$ model.

\subsection{$S$-wave states}

\subsubsection{$\Upsilon(11020)$}

Firstly, we give some discussions about a well-established resonance $\Upsilon(11020)$.
It is often assigned to the $\Upsilon(6S)$ state classified in the quark model.
In the present work, it is found that the $\Upsilon(11020)$ resonance may
have significant hadron continuum components, while the $b\bar{b}$-core component is only $\sim40~\%$.
The decay width is predicted to be
\begin{eqnarray}
\Gamma \simeq 16.5 \ \ \mathrm{MeV},
\end{eqnarray}
which is consistent with the PDG averaged value $\Gamma_{exp}=24^{+8}_{-6}$ MeV~\cite{ParticleDataGroup:2024cfk}.
Our results show that the $\Upsilon(11020)$ dominantly decays into the $\bar{B}B(1P_1)$ channel
with a branching fraction of $\sim 75\%$.
Thus, the $\Upsilon(11020)$ may provide a good source for looking for the missing
orbitally excited state $B(1P_1)(5656)$ predicted in the quark model~\cite{Ni:2023lvx}.
From Table~\ref{4S5S6S}, one finds that the decay rates of $\Upsilon(11020)$ into
the $\bar{B}B$, $\bar{B}_sB_s$, $\bar{B}B^*$, and $\bar{B}_sB_s^*$ channels are
predicted to be tiny, their branching fractions are in the order of $1\%$,
which may be a nature explanation why $\Upsilon(11020)$ has not been established
in these open bottom channels.


\subsubsection{$\Upsilon(10860)$}

For the $\Upsilon(10860)$ resonance, as the $\Upsilon(5S)$ assignment,
the decay properties cannot be well explained in theory, although the observed mass is consistent with the quark model predictions.
If taking the measured mass $M_{exp}=10885$ MeV, the width is predicted to be
\begin{eqnarray}
\Gamma \simeq 8 \ \ \mathrm{MeV},
\end{eqnarray}
which is notably narrower than the observed one $\Gamma_{exp} \simeq 37 \pm 4$ MeV~\cite{ParticleDataGroup:2024cfk}.
It should be mentioned that the $\bar{B}^*B(1^3P_0)$ channel may open.
By combining the predicted mass of $M = 5573$ MeV for the $B(1^3P_0)$ state in our previous work~\cite{Ni:2023lvx}, the mass threshold of the $\bar{B}^*B(1^3P_0)$ channel is predicted to be $\sim 10.9$ GeV, which is very close to the mass of $\Upsilon(10860)$.
If the $\bar{B}^*B(1^3P_0)$ channel opens, it can contribute a fairly large partial width of $\sim 30-40$ MeV due to its strong $S$-wave coupling to the $\Upsilon(5S)$.
Then, the total width of $\Upsilon(5S)$ can reach up to $\Gamma\sim 40-50$ MeV, which is consistent with the measured width of $\Upsilon(10860)$.
However, if the $\bar{B}^*B(1^3P_0)$ channel opens, the $\Upsilon(10860)$ should have a large decay rate ($\sim70\%$) into the  $\bar{B}^*B\pi$ channel via the cascade decay, which is inconsistent with the observation ($\sim7.3\%$) from the Belle Collaboration~\cite{Belle:2010hoy}.

On the other hand, if assigning $\Upsilon(10860)$ to the $\Upsilon(5S)$ state, there also
exists a puzzle in the decay modes.
According to our predictions, the $\Upsilon(5S)$ state mainly decays into the $\bar{B}_s^*B_s^*$, $\bar{B}_sB_s^*$ and $\bar{B}B^*$ channels if
the $\bar{B}^*B(1^3P_0)$ channel doesn't open.
However, the observation shows that the $\Upsilon(10860)$ resonance dominantly decays into the $\bar{B}^*B^*$, $\bar{B}^*_sB^*_s$, and $\bar{B}B^*$ channels with
branching fractions of $\sim 38 \%$, $\sim 18 \%$, and $\sim 14 \%$, respectively~\cite{ParticleDataGroup:2024cfk}.
There have been many studies on the decay properties of the $\Upsilon(5S)$ in the literature~\cite{Ferretti:2013vua, Godfrey:2015dia, Lu:2016mbb, Segovia:2016xqb, Wang:2018rjg, Li:2019qsg, Asghar:2023fvk, Ortega:2024rrv}.
In these works, the $\bar{B}B^*$ and $\bar{B}_s^*B_s^*$ are often predicted to be the main channels, while the decay rate into the $\bar{B}^*B^*$ channel is often small.

\begin{figure}
\epsfxsize=9.2 cm \epsfbox{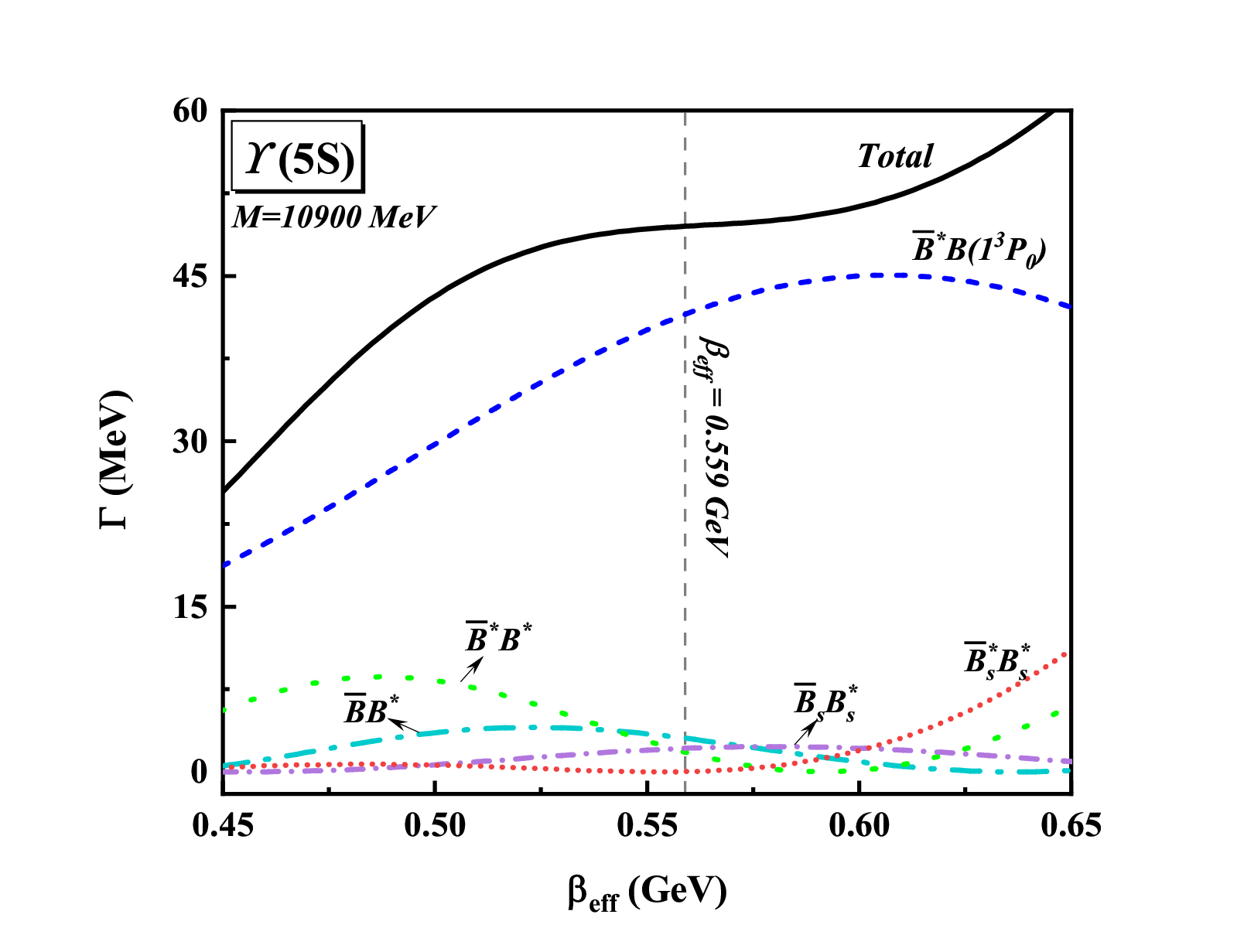}\vspace{-0.5 cm} \caption{The total and partial decay widths of $\Upsilon(5S)$ as a function of its wave function size parameter $\beta_{eff}$. The $\bar{B}B$ and $\bar{B}_sB_s$ channels are neglected due to their tiny contributions.}
\label{Upsilon5SBetaRunning}
\end{figure}

The discrepancy between theoretical predictions and observations may arise from the uncertainty of the wave function of the $\Upsilon(5S)$ predicted in the quark model.
To see the wave function dependency of our predictions, we first fit the numerical wave function obtained from the potential model by using a simple harmonic oscillator (SHO) form, and obtain the size parameter $\beta_{eff} = 0.559$ GeV.
Then, we calculate the decay properties with the SHO wave function by
varying the size parameter around the region of $\beta_{eff} = 0.559$ GeV estimated from the potential model.
When calculating the decay width, we adopted a mass of $M = 10900$ MeV, which is slightly above the $\bar{B}^*B(1^3P_0)$ threshold, to include this channel contribution.
Our results have been shown in Fig.~\ref{Upsilon5SBetaRunning}.
It is seen that the partial widths of $\bar{B}^*B^*$, $\bar{B}B^*$, $\bar{B}_s^*B_s^*$, and $\bar{B}_sB_s^*$ are rater sensitive to the details of the wave function.
If taking $\beta_{eff} = 0.500$ GeV, the $\bar{B}^*B^*$ and $\bar{B}B^*$ are the main decay channels of the $\Upsilon(5S)$, which seems to favor the observation.
However, in this case the decay rates of $\Upsilon(5S)\to \bar{B}_s^*B_s^*,\bar{B}_sB_s^*$ are too small to be comparable with the data.

Finally, it should be mentioned that the $\Upsilon(10860)$ may be partly contributed by some exotic components as well.
For example, in Ref.~\cite{Ali:2009pi} the analysis of the data of $e^+e^- \to b\bar{b}$ cross section from the BaBar Collaboration~\cite{BaBar:2008cmq} may indicate two overlapping resonances in the $\Upsilon(5S)$ mass region.
One corresponds to the $\Upsilon(5S)$ state, while the other may correspond to
a bottomonium-like tetraquark state.
To uncover the puzzles of the $\Upsilon(10860)$ resonance,
more observations are waiting to be carried out in future experiments.

\subsubsection{higher $S$-wave states}

For the other missing $S$-wave states, $n^1S_0$ ($n=5,6,7$) and $7^3S_1$, our predicted decay properties have been given in Tables~\ref{4S5S6S} and \ref{7S}.
These states may be very narrow states with a width of $\Gamma\simeq1-20$ MeV.
The $5^1S_0$ state may have a mass of $M=10840$ MeV, and dominantly decays into the $\bar{B}_s^*B_s^*$, $\bar{B}^*B^*$, and $\bar{B}_sB_s^*$ channels with branching fractions $\sim64~\%$, $\sim25~\%$, $\sim10~\%$, respectively.
The $7^1S_0$ and $7^3S_1$ states have a nearly degenerated mass of $M=11255$ MeV
and a comparable width of $\Gamma\simeq 22$ MeV.
They mainly decay into the channels containing excited $B$ meson states, while their decay rates into the $\bar{B}^{(*)}B^{(*)}$ and $\bar{B}^{(*)}_sB^{(*)}_s$ are very tiny.


\subsection{$P$-wave states}

Since the $3P$ states $\chi_{b1}(3P)$ and $\chi_{b2}(3P)$ have been established
in experiments~\cite{ParticleDataGroup:2024cfk}, the higher $4P$ states may have large potentials to be observed in future experiments.
According to the quark model predictions, their masses scatter in the range of $10.77-10.79$ GeV, thus, some OZI-allowed two-body strong decay channels open.
We evaluate the strong decay properties of the $4P$ states, and list the results in Table~\ref{3P4P5P}.
It is found that the $4P$ $b\bar{b}$ states may be narrow states.
For the $0^+$ state $\chi_{b0}(4P)$, the width is predicted to be $\Gamma\simeq 10$ MeV.
This state mainly decays into $\bar{B}B$ and $\bar{B}^*B^*$ channels with branching fractions $\sim57~\%$ and $\sim41~\%$, respectively.
While for the two $J^P=1^+$ states, $\chi_{b1}(4P)$ and $h_b(4P)$, the widths are predicted to
be only a few MeV.
They may dominantly decay into the $\bar{B}B^*$ channel.
The $2^+$ state $\chi_{b2}(4P)$ has a narrow width of $\Gamma\simeq 11$ MeV.
Besides the dominant decay channel $\bar{B}B^*$, it also has sizable decay rates into the $\bar{B}B$, $\bar{B}_sB_s$ and $\bar{B}^*B^*$ channels with a comparable branching fraction of $\sim20~\%$.
To establish the $4P$ states, the decay channels $\bar{B}B$, $\bar{B}B^*$, $\bar{B}^*B^*$ and $\bar{B}_sB_s$ are worth observing in future experiments.

For the higher $5P$ and $6P$ excitations, the masses are predicted in the ranges of $10.96-10.98$ GeV and $11.20-11.22$ GeV, respectively.
By combining the predicted masses and wave functions, we further evaluate the strong decay properties.
Our results are given in Tables~\ref{3P4P5P} and \ref{6P}.
The $5P$ may be very narrow states with a width of $\Gamma\simeq 6-20$ MeV, while the $6P$ may be relatively broader states with a width of $\Gamma\simeq 40-60$ MeV.
The $5P$ and $6P$ states couple weakly to the common channels $\bar{B}^{(*)}B^{(*)}$ and $\bar{B}^{(*)}_sB^{(*)}_s$, while often couple strongly to the channels who contain an excited $B$ meson.
For example, the $\chi_{b2}(6P)$ state weakly couples to the $\bar{B}^{(*)}B^{(*)}$
and $\bar{B}^{(*)}_sB^{(*)}_s$ channels, while strongly couples to the $\bar{B}^*B(2^3S_1)$ channel.
To know more details of the strong decay properties of the $5P$ and $6P$ states, one can see Tables~\ref{3P4P5P} and \ref{6P}.

\subsection{$D$-wave states}

From our model, the mass range of the $3D$ states is $10.67-10.72$ GeV.
They have phase spaces for decaying into the low-lying OZI-allowed channels, such as $\bar{B}B$, $\bar{B}B^*$, $\bar{B}^*B^*$, and so on.
The strong decay properties are given in Table~\ref{3D4D5D}.
For one case in $J^P=1^-$ and two cases with different total quark spins in $J^P=2^-$, there is only one solution for each case.
The predicted widths of $\Upsilon_{1}(3D)(10670)$, $\Upsilon_{2}(3D)(10685)$ and $\eta_{b2}(3D)(10691)$
are $\Gamma \simeq 31,~81$ and $53$ MeV, respectively.
They mainly decay into both the $\bar{B}B^*$ and $\bar{B}^*B^*$ channels with
a comparable ratio $\sim1:2$.
On the other hand, there are two physical solutions with $J^P=3^-$, $\Upsilon_3(10650)$ and $\Upsilon_3(10718)$ obtained in the unquenched quark model.
The lower mass resonance $\Upsilon_3(10650)$ may have a narrow width of $\Gamma\simeq 19$ MeV,
and mainly decays into the $\bar{B}B$ ($\sim 45\%$) and $\bar{B}B^*$ ($\sim 55\%$) final states, while the higher mass resonance $\Upsilon_3(10718)$ has a relatively broader width of
$\Gamma\simeq 63$ MeV, and dominantly decays into the $\bar{B}B^*$ ($\sim 40\%$) and $\bar{B}^*B^*$ ($\sim 53\%$) final states.
Thus, to establish the $3D$ states, the $\bar{B}B$, $\bar{B}B^*$ and $\bar{B}^*B^*$
channels provides a nice place for observing in future experiments.

The masses of the $4D$ states are estimated to be in the range of $10.91-10.93$ GeV, which is
about $200$ MeV above the $3D$ states.
With our predicted masses, the strong decay properties are analyzed, as listed in Table~\ref{3D4D5D}.
Here it shows that the $4D$ states are narrow resonances with a width of $\Gamma\simeq 10-20$ MeV.
These states dominantly decay into the low-lying open bottom channels, $\bar{B}^{(*)}B^{(*)}$ and $\bar{B}^{(*)}_s B^{(*)}_s$.
For example, the $\Upsilon_{1}(4D)$ state may dominantly decay into $\bar{B}^*B^*$ and $\bar{B}^*_sB^*_s$ channels with a comparable decay rate of $\sim 35\%$, while also has
sizable decay rates into the $\bar{B}B$ ($\sim 8\%$) and $\bar{B}B^*$ ($\sim 10\%$) channels.
To establish the $4D$ states, the $\bar{B}B^*$, $\bar{B}^*B^*$, and $\bar{B}^*_sB^*_s$
channels are worth observing in future experiments, please note it is a bit different from the case of $3D$ states.

For the higher $5D$ and $6D$ states, the masses are predicted to be in the ranges of $11.06-11.10$ GeV and $11.30-11.31$ GeV, respectively.
By using the predicted masses, we further study the strong decay properties of these $5D$ and $6D$ states, the results are given in Tables~\ref{3D4D5D} and \ref{6D}, respectively.
The widths of these state are significant larger than that of $4D$ states.
These higher $5D$ and $6D$ states mainly decay into the open-bottom channels containing a $B$-meson excitation, while their decay rates into the $\bar{B}^{(*)}B^{(*)}$ and $\bar{B}^{(*)}_sB^{(*)}_s$ final states are tiny.
Thus, the $5D$ and $6D$ states may be difficult to be established in the $\bar{B}^{(*)}B^{(*)}$
and $\bar{B}^{(*)}_sB^{(*)}_s$ final states.
For example, to look for the $\Upsilon_1(5D)$ state one should select its dominant decay channel $\bar{B}^*B_1(5721)$.

\section{$S$-$D$ mixing}\label{SDMix}

The $n^3S_1$ and $m^3D_1$ ($n=m+1$) states with $J^{PC}=1^{--}$ usually have a similar mass.
A sizable mixing between the $n^3S_1$ and $m^3D_1$ should be induced by the tensor force and/or the intermediate hadronic loops.
Considering the $S$-$D$ mixing, the physical states $\Upsilon'(nS)$ and $\Upsilon'_1(mD)$, which are dominated by the $n^3S_1$ and $m^3D_1$ components, respectively, are defined by
\begin{eqnarray}\label{sdmix}
\left(\begin{array}{c}\Upsilon'(nS)  \cr \Upsilon'_1(mD)
\end{array}\right)
=\left(\begin{array}{cc}~~ \cos\phi_S & \sin\phi_S\cr -\sin\phi_D & \cos\phi_D
\end{array}\right)
\left(\begin{array}{c} \Upsilon(nS)  \cr  \Upsilon_1(mD)
\end{array}\right),
\end{eqnarray}
where $\phi_{S/D}$ is the mixing angle.
In the literature, the $S$-$D$ mixing are often artificially introduced to explain the dielectron decay widths of the vector resonances.
However, the underlying dynamic mechanisms for the $S$-$D$ mixing are often overlooked.
Here, we will introduce the mixing mechanisms based on this unquenched quark model.

\subsubsection{Mixing induced by the tensor force}

Firstly, we consider the $S$-$D$ mixing induced by the following tensor force,
\begin{eqnarray}\label{Vt}
V_{T}(r)=\frac{4}{3}\frac{\alpha_s}{m_{b} m_{\bar{b}}}\frac{1}{r^3}\left(\frac{3\mathbf{S}_{q}\cdot \mathbf{r}\,\mathbf{S}_{{\bar{q}}}\cdot \mathbf{r}}{r^2}-\mathbf{S}_{q}\cdot\mathbf{S}_{{\bar{q}}}\right).
\end{eqnarray}
The mixing angle can be determined by
\begin{eqnarray}\label{Mix matrix equationa}
&&\left(\begin{array}{cc}
M_{nS} & \tilde{V}_T \\
\tilde{V}_T & M_{mD}
\end{array}\right)
\left(\begin{array}{c}
a_{S} \\
a_{D}
\end{array}\right)
=M\left(\begin{array}{c}
a_{S} \\
a_{D}
\end{array}\right),
\end{eqnarray}
where the $M_{nS}$ and $M_{mD}$ are the bare masses of the $\Upsilon(nS)$ and $\Upsilon_1(mD)$ states, respectively.
The values of these masses have been determined within the potential model.
The non-diagonal element $\tilde{V}_T$ is determined by $\tilde{V}_T=\langle n^3S_1|V_T(r)|m^3D_1\rangle$.
By solving Eq.~(\ref{Mix matrix equationa}), one can obtain
the physical masses and the eigenvectors of the mixed states.
Finally, the mixing angle is given by $\phi_S= \phi_D= \arctan(a_{D}/a_{S})$.
Our calculations obtain a tiny mixing angle for this mechanism, $\phi_{S/D} \simeq -0.2^{\circ}$, for the $S$-$D$ mixing induced by the tensor term, which is consistent with the predictions in Refs.~\cite{Badalian:2009bu, Lu:2016mbb}.


\subsubsection{Mixing induced by hadronic loops}

Then, we further consider the $S$-$D$ mixing induced by the intermediate hadronic loops.
The mixing angle can be approximately determined by
\begin{eqnarray}\label{Mix matrix equation}
\left(\begin{array}{cc}
M_{S}+\Delta M_{S}(M) &\Delta M_{SD}(M) \\
\Delta M_{DS}(M) & M_{D}+\Delta M_{D}(M)
\end{array}\right)
\left(\begin{array}{c}
a_{S} \\
a_{D}
\end{array}\right)
=M\left(\begin{array}{c}
a_{S} \\
a_{D}
\end{array}\right). \nonumber \\
\end{eqnarray}
%
The mass shifts $\Delta M_{S}(M)$ and $\Delta M_{D}(M)$ of bare states $\Upsilon(nS)$ and $\Upsilon_1(mD)$ are estimated by Eq.~(\ref{ReDelta MM0}), respectively, while the off-diagonal elements $\Delta M_{SD}(M)$ and $\Delta M_{SD}(M)$ are determined by
\begin{eqnarray}\label{DeltaA1A2A12}
&&\Delta M_{SD}(M) = \Delta M^{*}_{DS}(M) \nonumber\\
&=&\sum_{BC} \mathcal{P}
\int
\frac{(M_{0}-M)\overline{\mathcal{M}_{S \rightarrow B C}^*(\boldsymbol{p}) \mathcal{M}_{D \rightarrow B C}(\boldsymbol{p}) }}{(M-E_{B C})(M_{0}-E_{B C})} p^{2} dp.
\end{eqnarray}
The Eq.(\ref{Mix matrix equation}) can be transformed into an eigenvalue problem, i.e.,
\begin{eqnarray}\label{detSD}
\mathrm{det}\left|
\begin{matrix}
{M_{S}+\Delta M_{S}(M)-M}
&{\Delta M_{SD}(M)}\\[1pt]
{\Delta M_{DS}(M)}
&{M_{D}+\Delta M_{D}(M)-M}\\[1pt] \end{matrix}\right|=0. \nonumber\\
\end{eqnarray}
With the constraint of Eq.~(\ref{M=MA+Delta M}), one can determine the physical masses, $M$, for the $\Upsilon'(nS)$ and $\Upsilon'_1(mD)$, and their corresponding eigenvector by diagonalizing Eq.(\ref{detSD}).
The mixing angles $\phi_{S}$ and $\phi_{D}$ for the $\Upsilon'(nS)$ and $\Upsilon'_1(mD)$
are given by $\phi_{S} = \arctan(a_{D}/a_{S})$ and $\phi_{D}=-\arctan(a_{S}/a_{D})$.

The obtained $S$-$D$ mixing angles induced by the hadronic loops are listed in Table \ref{SD mixing angle(hadron loop)}.
It is interestingly found that there are a fairly large mixing angle $\phi_D\simeq-(14\sim18)^\circ$ for the $\Upsilon'_1(mD)$ ($m=3,5,6$) states.
In other words, in the $b\bar{b}$ core of $\Upsilon'_1(mD)$ ($m=3,5,6$), besides the dominant
$nD$-wave component, there is a sizeable $\Upsilon(nS)$ ($n=m+1$) component due to the coupled-channel effects.
However, for the $S$-wave dominant state $\Upsilon'(nS)$ ($n=4,5,6,7$), the $D$-wave $b\bar{b}$ core component is small, with the mixing angle of only a few degrees.

In a previous work of our group~\cite{Li:2019qsg}, the $\Upsilon(10753)$ is suggested to be a
$4D$-wave dominant state via the $\Upsilon_1(4D)$-$\Upsilon(5S)$ mixing if the mixing
angle reaches up to a fairly large value $\phi_D=20^\circ-30^\circ$.
However, with the hadronic loops and tensor forces we cannot explain such a fairly large mixing angle.
There is only a small mixing angle, $\phi_D\simeq4^\circ$, induced by the hadronic loops.

Finally, it should be mentioned that the strong decay properties of the $\Upsilon_1(4D)$ and $\Upsilon_1(5D)$ may be significantly changed by the $S$-$D$ mixing.
Considering this mixing effects, the physical state $\Upsilon'_1 (3D) $ has a broad width
\begin{eqnarray}
\Gamma[\Upsilon'_1 (3D)] &=& 57 \ \ \mathrm{MeV},
\end{eqnarray}
which is nearly saturated by the $\bar{B}^{*} B^{*}$ channel ($\sim 92~\% $).
The decay width is about a factor of $2$ broader than the case without mixing.
While for the mixed state $\Upsilon'_1(5D)$, the decay width is predicted to be
\begin{eqnarray}
\Gamma[\Upsilon_1'(5D)] &\simeq& 64 \ \ \mathrm{MeV},
\end{eqnarray}
and it mainly decays into the $\bar{B}^{*}B_2^{*}(5747)$ ($\sim 53~\%$), $\bar{B}^{*}B_2(5721)$ ($\sim 24~\%$), and $\bar{B}B(2^1S_0)$ ($\sim 8~\%$) final states.
The $S$-$D$ mixing significantly increases the decay width of $\Upsilon_1(5D)$.

\section{Dielectron decay of vector states}\label{aab}

Taking into account QCD radiative corrections with the Van Royen-Weisskopf formula~\cite{VanRoyen:1967nq}, the dielectron decay widths of the vector $b\bar{b}$ states are given by~\cite{Barbieri:1975ki, Novikov:1977dq, Barbieri:1979be}
\begin{eqnarray}\label{Gammaee}
	\Gamma_{ee}[\Upsilon(nS)] &=&\frac{4\alpha_0^2e_b^2}{M_{nS}^2}\left|R_{nS}(0)\right|^2\left(1-\frac{16\alpha'_s}{3\pi}\right),  \\
\label{Gammaee2}
	\Gamma_{ee}[\Upsilon_1(nD)] &=&\frac{4\alpha_0^2e_b^2}{M_{nD}^2}\left|\frac5{2\sqrt{2}m_b^2}R_{nD}^{\prime\prime}(0)\right|^2\left(1-\frac{16\alpha'_s}{3\pi}\right),
\end{eqnarray}
where $M_{nS}$ and $M_{nD}$ are the masses of the $\Upsilon(nS)$ and $\Upsilon_1(nD)$, respectively, $e_{b}=-1/3$ is the charge of $b$ quark, $\alpha_0 \simeq 1/137$ is the fine-structure constant, and $\alpha'_s$ is strong coupling constant.
Here, we take $\alpha'_s=0.18$ as adopted in Refs.~\cite{Li:2009nr,Wang:2018rjg,Li:2021jjt}.
$R_{nS}(0)$ is the wave function of $\Upsilon(nS)$ at the zero point, while $R_{nD}^{\prime\prime}(0)$
is the second derivative of the wave function of $\Upsilon_1(nD)$ at the zero point.

It should be emphasized that the dielectron decay widths of some vector resonances
may be significantly affected by the following two aspects:
i) The continuum components of a vector state at the long range should significantly suppress the dielectron width.
As we know, the dielectron decay width is contributed by the zero point wave function, which
is governed by the short range $q\bar{q}$ core component rather than the long range continuum components.
The anomalously small dielectron width of $\Upsilon(10580)$ may be due to the suppression of the
significant continuum components~\cite{Lu:2016mbb}.
ii) The $S$-$D$ mixing induced by the intermediate hadronic loops can affect the dielectron decay width as well.
Thus, in the unquenched quark model framework, and including the $S$-$D$ mixing effects, the dielectron decay widths for the physical states are given by
\begin{eqnarray}\label{GammaeeSDmixing}
&&\Gamma_{ee}[\Upsilon'(nS)]
=\frac{4\alpha_0^2e_b^2}{M_{nS}^2}\left(1-\frac{16\alpha_s}{3\pi}\right) \\ \nonumber
&&~~~~~~~~~~~~
\left|c_{nS}R_{nS}(0)\cos \phi + \frac5{2\sqrt{2}m_b^2}c_{mD}R_{mD}^{\prime\prime}(0) \sin \phi \right|^2, \label{GammaeeSDmixing2}\\
&&\Gamma_{ee}[\Upsilon'_1(mD)]
=\frac{4\alpha_0^2e_b^2}{M_{mD}^2}\left(1-\frac{16\alpha_s}{3\pi}\right)
\\ \nonumber
&&~~~~~~~~~~~~
\left|-c_{nS}R_{nS}(0)\sin \phi + \frac5{2\sqrt{2}m_b^2}c_{mD}R_{mD}^{\prime\prime}(0) \cos \phi \right|^2,
\end{eqnarray}
where $c_{nS}$ and $c_{mD}$ are the probability amplitudes of the $b\bar{b}$ core components for the $\Upsilon(nS)$ and $\Upsilon_1(mD)$ states, respectively.

According to Eqs.(\ref{GammaeeSDmixing}) and (\ref{GammaeeSDmixing2}), we evaluate the
dielectron decay widths for the vector $\Upsilon(nS/nD)$ states.
Our results are listed in Column 9 of Table~\ref{ee Decay Width and Ratio}, labeled as Case A.
It is found that our predictions for the $\Upsilon(nS)$ ($n=1-6$) are in good agreement with the observations~\cite{ParticleDataGroup:2024cfk}.
For a comparison, we also give the quenched quark model results predicted with Eqs.(\ref{Gammaee}) and (\ref{Gammaee2}) in Column 11 of Table~\ref{ee Decay Width and Ratio}, labeled as Case C.
One can see that neglecting the affects of the continuum components, the dielectron decay widths
for the high-lying resonances $\Upsilon(10580)$, $\Upsilon(10860)$, and $\Upsilon(11020)$
seem to be notably overestimated in the quenched quark model framework.
Moreover, to examine the $S$-$D$ mixing effects, we further evaluate the dielectron decay widths without mixing, i.e., $\phi=0$ for Eqs.(\ref{GammaeeSDmixing}) and (\ref{GammaeeSDmixing2}), the results are listed in Column 10 of Table~\ref{ee Decay Width and Ratio},
labeled as Case B.
It is seen that the $S$-$D$ mixing has small effects on the dielectron decay widths
of the $S$-wave states, while it has notably effects on those of
the high-lying $D$-wave states $\Upsilon_1(nD)$ $(n=3-6)$.

Finally, it should be mentioned that recently, the Belle II Collaboration
reported their new measurements of the $e^{+}e^{-} \to \bar{B}^*B^*$ cross sections~\cite{Belle-II:2024niz}.
There seems to be a rapid increase in the cross section around the $\bar{B}^*B^*$ threshold,
which may be caused by the missing $\Upsilon'_1(3D)$ state predicted in the quark model.
The reasons are given as follows.
(i) The predicted mass, $10670$ MeV, for the $\Upsilon'_1(3D)$ state is consistent with that of the observed structure in the cross section.
(ii) For the $\Upsilon'_1(3D)$, there may be enough production rates via $e^{+}e^{-}$ annihilation at Belle/Belle II.
According to our prediction, the $\Upsilon'_1(3D)$ may have a relatively large dielectron decay width, $\Gamma_{ee}\simeq 0.028$ keV, due to the significant $S$-$D$ mixing induced by the hadronic loops.
(iii) The $\Upsilon'_1(3D)$ has a relatively narrow width of $\Gamma\simeq 57$ MeV, and dominantly decays into the $\bar{B}^*B^*$ final state with a branching fraction of $\sim 92~\%$.
This is also consistent with the Belle II observations. 

\begin{figure*}
\centering \epsfxsize=17.0 cm \epsfbox{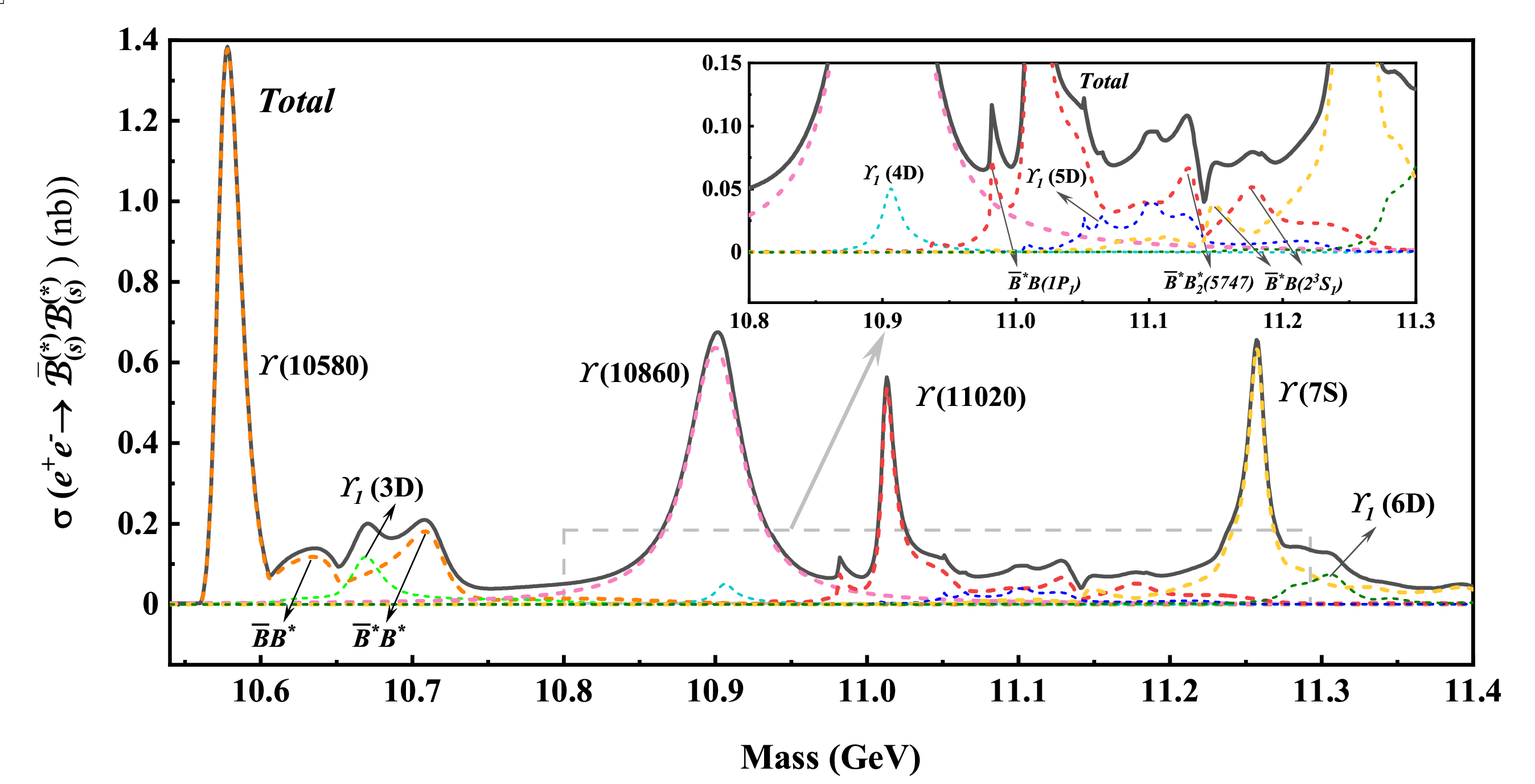} \vspace{-0.2 cm} \caption{Estimation of the cross section of the $e^+e^-\to \bar{\mathcal{B}}^{(*)}_{(s)}\mathcal{B}^{(*)}_{(s)}$ process (dashed curve).
It should be emphasized that label $\mathcal{B}^{(*)}_{(s)}$ includes both the ground-state $B^{(*)}_{(s)}$ mesons and their excited states.
For clarity, the contributions from the
vector resonances and some important threshold effects of coupled channels are labeled in the figure as well.}
\label{bbbarCS}
\end{figure*}

\section{The roles of vector resonances in $e^+e^-\to \bar{\mathcal{B}}^{(*)}_{(s)}\mathcal{B}^{(*)}_{(s)}$ cross section}\label{crosss}

In experiments, the production of $b\bar{b}$ vector resonances mainly depends on the $e^{+}e^{-}$ annihilation reaction $e^{+}e^{-}\to b\bar{b}$.
The high-lying $b\bar{b}$ vector resonances are most likely to be reconstructed in the open bottom meson pair channels.
Thus, the study of the cross section for the process $e^+e^- \to b\bar{b} \to\bar{\mathcal{B}}^{(*)}_{(s)}\mathcal{B}^{(*)}_{(s)}$ (where $\mathcal{B}^{(*)}$ and $\mathcal{B}^{(*)}_{s}$ denote both the ground-state $B^{(*)}$ and $B^{(*)}_{s}$ mesons, as well as their excitations) is essential for understanding the properties of $b\bar{b}$ vector resonances.
If we neglect the contribution from the no-resonance backgrounds, the total cross-section for the $e^{+}e^{-} \to \bar{\mathcal{B}}^{(*)}_{(s)}\mathcal{B}^{(*)}_{(s)}$ above the $B\bar{B}$ threshold could be roughly estimated by~\cite{Konigsmann:1987yb, Klempt:2004yz}
\begin{eqnarray}\label{CRS}
\sigma(e^{+}e^{-} \to \bar{\mathcal{B}}^{(*)}_{(s)}\mathcal{B}^{(*)}_{(s)}) &\simeq& \sum_R\frac{3\pi}{M^2} \frac{\Gamma_{e^{+}e^{-}}\Gamma_{R}}{(M-M_{R})^2 + \Gamma^2_{R}/4},
\end{eqnarray}
where $M_R$ and $\Gamma_{R}$ stand for the mass and width of the intermediate vector $b\bar{b}$ resonances, $M$ is the center-of-mass energy.

By using Eq.(\ref{CRS}), we calculate the cross section of $e^{+}e^{-} \to \bar{\mathcal{B}}^{(*)}_{(s)}\mathcal{B}^{(*)}_{(s)}$ in the center-of-mass energy range of $(10.55,11.35)$ GeV.
In this energy range, 8 vector states, $\Upsilon(nS)$ ($n=4,5,6,7$) and $\Upsilon_1(mD)$ ($m=3,4,5,6$), are included according to our quark model prediction.
Our results are plotted in Fig.~\ref{bbbarCS}.
It should be mentioned that similar analysis was carried out recently by Ortega~\emph{et al.}~\cite{Ortega:2024rrv}.
In Fig.~\ref{bbbarCS}, one can see four prominent peak structures, which contributed by four $S$-wave states $\Upsilon(4S)$, $\Upsilon(5S)$, $\Upsilon(6S)$, and $\Upsilon(7S)$, respectively.
The $\Upsilon(4S)$, $\Upsilon(5S)$, and $\Upsilon(6S)$ have been established in experiments, which should correspond to the $\Upsilon(10580)$, $\Upsilon(10860)$, and $\Upsilon(11020)$, respectively.
However, the $\Upsilon(7S)$ with a mass of $\sim11260$ MeV is still missing, this state may be hard to observe in the $\bar{B}^{(*)}_{(s)}B^{(*)}_{(s)}$ channels due to their tiny decay rates, however it may be observed in the $\bar{B}^{(*)}B_1(5721)$ channels.

For the $3D$ state, by including $S$-$D$ mixing effect, the physical state $\Upsilon_1'(3D)(10670)$ has a relatively clear role in the cross section distribution, which suggests to be observed by the Belle II Collaboration, recently~\cite{Belle-II:2024niz}.
However, it is notable that the high statistic data will be required to extract the clean signal of the  $\Upsilon_1'(3D)(10670)$, since it would suffer from the interference due to the threshold effects of  the $\bar{B}B^*$, $\bar{B}^*B^*$ channels.
On the contrary, the high-lying $D$-wave states, $\Upsilon_1(mD)$ ($m=4,5,6$), do not show clear peak in the total cross section distribution.

At last, let us make further discussion about the threshold effects of the coupled channels.
As we know, the threshold effects can make bump or cusp around the threshold.
Here, as shown in Fig.~\ref{bbbarCS}, there are two bumps due to the threshold effects of $\bar{B}B^*$ and $\bar{B}^*B^*$ channels around $10.63$ and $10.71$ GeV, respectively.
Actually, the bump near $10.71$ GeV in our prediction provides a possible explanation for the bump, known as $\Upsilon(10753)$ resonance, observed at Belle/Belle II~\cite{Belle:2019cbt, Belle-II:2024mjm}.
As discussed before, $\Upsilon(10753)$ resonance can not be assigned as $\Upsilon_1(3D)$ nor $\Upsilon_1(5S)$.
Even a mixed state via $\Upsilon(4S)$-$\Upsilon_1(3D)$ or $\Upsilon(5S)$-$\Upsilon_1(4D)$ is ruled out in our current model due to the small mixing angle.
Now, it is clear in our model that such bump is just the threshold effect of $\bar{B}^*B^*$ coupling with $\Upsilon(4S)$, rather than a genuine state.
Similar phenomena are also reported by Ortega \emph{et al.} in their recent work~\cite{Ortega:2024rrv}.
Of course, it still need more evidence to confirm it.
The interpolations of $\Upsilon(10753)$ as a hybrid~\cite{Brambilla:2019esw, TarrusCastella:2021pld} or a tetraquark state~\cite{Ali:2019okl, Bicudo:2020qhp, Bicudo:2022ihz, Wang:2019veq} are also possible.

\section{Summary}\label{SUM}

In this work, the bottomonium spectrum and the OZI-allowed two-body strong decay widths are systematically studied within an unquenched quark model.
Furthermore, we study the $S$-$D$ mixing and dielectron decay properties in the same framework to extract more detail of these vector states.
Moreover, by combining the spectral density functions and dielectron decay widths,
we discuss the roles of the vector $b\bar{b}$ states in the cross-section of the $e^{+}e^{-}$ annihilation reaction.
Some key results from this study are summarized as follows:

\begin{itemize}
\item The mass shifts of the bare states due to the unquenched coupled-channel effects are estimated to be $\mathcal{O}(10)~\mathrm{MeV}$.
Including the coupled-channel effects, the theoretical description of the masses for the high-lying resonances, such as $\Upsilon(10580)$, $\Upsilon(11020)$, $\chi_{b1,2}(3P)$, is systematically improved.

\item Most of the high-lying resonances may contain significant non-$b\bar{b}$ components.
For example, for the $\Upsilon(11020)$ resonance the non-$b\bar{b}$ component may reach up to $\sim 60\%$.

\item The intermediate hadronic loops can cause a significant $S$-$D$ mixing for the high-lying vector states.
The $D$-wave states, $\Upsilon_1(3D)$, $\Upsilon_1(5D)$, and $\Upsilon_1(6D)$ will significantly mix with their nearby $S$-wave states.
The mixing angles is estimated to be $\phi \simeq -(14-18)^\circ$.

\item  The dielectron widths of some high-lying vector resonances may be significantly
suppressed by their non-$b\bar{b}$ components due to the coupled-channel effects.
Including these effects, the better description of the dielectron widths of the $\Upsilon(10580)$, $\Upsilon(10860)$ and $\Upsilon(11020)$ is achieved in this work.

\item The threshold effects of open-bottom meson pairs can cause rich bump structures in the total cross section distribution of $e^{+}e^{-}\to b\bar{b}$.
There are complex structures around the c.m. energy region of $10.62-10.75$ GeV, which are mainly contributed by the $\Upsilon_1(3D)$ resonance together with the threshold effects of
the $\bar{B}B^*$ and $\bar{B}^*B^*$.
In current model, we do not favor existing the $\Upsilon(10753)$ resonance, while the bump in the $e^{+}e^{-}\to b\bar{b}$ around $10.71$ GeV is just due to the threshold effect of $\bar{B}^*B^*$.

\end{itemize}

\begin{table}
	\begin{center}
		\caption{The mass shifts and strong decay width of
each channel for the $4S$-, $5S$-, and $6S$-wave bottomonium states, in units of MeV. For the forbidden/unopened channels, the absence values are labeled with ``--''. The mass shifts labeled with bold numbers are contributed
by the nearby virtual channels whose mass thresholds are slightly
above the bare state. The experimental data is taken from PDG~\cite{ParticleDataGroup:2024cfk}. }
		\label{4S5S6S}
		\tabcolsep=0.02cm
		\renewcommand\arraystretch{1.1}
		\scalebox{1.0}{
\begin{tabular}{ccccccccccccccccccccc}
				\bottomrule[1.0pt]\bottomrule[1.0pt]
&\multicolumn{2}{c}{$4^1S_0$}  &  &\multicolumn{2}{c}{$4^3S_1$}
				  \\
				\cline{2-3}\cline{5-6}
				 Channel                  &$\Delta M~[10608]$   &$\Gamma~[10571]$     &   &$\Delta M~[10619]$  &$\Gamma~[as~\Upsilon(10580)]$         \\
				\bottomrule[0.4pt]
				$\bar{B}B$               &$-$                  &$-$           &   &$-6.4$               &$20.5$      \\
				$\bar{B}_sB_s$           &$-$                  &$-$           &   &$-$                  &$-$           \\
				$\bar{B}B^*$             &$-22.0$              &$-$           &   &$-15.8$              &$-$           \\
				$\bar{B}^*B^*$           &$\mathbf{-15.2}$     &$-$           &   &$\mathbf{-17.9}$     &$-$          \\
				\emph{Total}             &$-37.2$              &$-$           &   &$-40.1$              &$20.5$    \\
				$M_{th},\Gamma_{th}$    &\multicolumn{2}{c}{$10571,-$}    &   &\multicolumn{2}{c}{$10579,20.5$}   \\
\end{tabular}}
	\end{center}
\end{table}

\begin{table}
 	\begin{center}
\caption{The mass shifts and strong decay width of
each channel for the $7S$-wave bottomonium states, in units of MeV. The caption is the same as that of Table~\ref{4S5S6S}. }
 		\label{7S}
 		\tabcolsep=0.06cm
 		\renewcommand\arraystretch{1.1}
 		\scalebox{1.0}{
 			\begin{tabular}{ccccccccccccccccccccc}
 				\bottomrule[1.0pt]\bottomrule[1.0pt]
 				&\multicolumn{2}{c}{$7^1S_0$}  &  &\multicolumn{2}{c}{$7^3S_1$}   \\
 				\cline{2-3}\cline{5-6}
 				Channel   &$\Delta M~[11262]$      &$\Gamma~[11254]$        &    &$\Delta M~[11269]$     &$\Gamma~[11259]$         \\
 				\bottomrule[0.4pt]
 				$\bar{B}B$                        &$-$                  &$-$              &     &$+0.01$              &$0.01  $            \\
 				$\bar{B}_sB_s$                    &$-$                  &$-$              &     &$\sim0$              &$\sim0   $            \\
 				$\bar{B}B^*$                      &$+0.1 $              &$0.1  $          &     &$+0.05$              &$0.08  $            \\
 				$\bar{B}_sB_s^*$                  &$+0.02$              &$\sim0$          &     &$\sim0$              &$\sim0   $            \\
 				$\bar{B}^*B^*$                    &$+0.06$              &$0.2  $          &     &$+0.01$              &$0.2  $            \\
 				$\bar{B}_s^*B_s^*$                &$+0.03$              &$0.03 $          &     &$+0.03$              &$0.02  $             \\
\end{tabular}}
 	\end{center}
 \end{table}

\begin{table*}
	\begin{center}
		\caption{The mass shifts and strong decay width of
each channel for the $3P$-, $4P$- and $5P$-wave bottomonium states, in units of MeV. The caption is the same as that of Table~\ref{4S5S6S}.}
		\label{3P4P5P}
		\tabcolsep=0.18cm
		\renewcommand\arraystretch{1.0}
		\scalebox{1.0}{
			\begin{tabular}{cccccccccccccccccccccccc}
				\bottomrule[1.0pt]\bottomrule[1.0pt]
				& &\multicolumn{2}{c}{$3^3P_0$}  &  &\multicolumn{2}{c}{$3^3P_2$}  &
				&\multicolumn{2}{c}{$3^1P_1$}   &   &\multicolumn{2}{c}{$3^3P_1$} \\
				\cline{3-4}\cline{6-7}\cline{9-10}\cline{12-13}
				Channel			&	&$\Delta M$~[10514]  &$\Gamma$~[10498]  &   &$\Delta M$~[10550]  &$\Gamma$~[10523] &
				&$\Delta M$~[10538]  &$\Gamma$~[10521]  &   &$\Delta M$~[10536]  &$\Gamma$~[10514]     \\
				\bottomrule[0.4pt]
				$\bar{B}B$          &         &$\mathbf{-15.5}$       &$-$          &   &$\mathbf{-3.5}$              &$-$         &   &$-$                  &$-$        &   &$-$                  &$-$           \\
				$\bar{B}B^*$        &         &$-$                    &$-$          &   &$\mathbf{-6.9}$              &$-$         &   &$\mathbf{-17.3}$     &$-$        &   &$\mathbf{-21.9}$     &$-$            \\
				$\bar{B}^*B^*$      &         &$-$                    &$-$          &   &$\mathbf{-16.5}$             &$-$         &   &$-$                  &$-$        &   &$-$                  &$-$          \\
				\emph{Total}        &         &$-15.5$                &$-$          &   &$-26.9$             &$-$         &   &$-17.3$              &$-$        &   &$-21.9$              &$-$        \\
				$M_{th},\Gamma_{th}$     &  &\multicolumn{2}{c}{$10498,-$}               &  &\multicolumn{2}{c}{$10523,-$}  &   &\multicolumn{2}{c}{$10521,-$}    &   &\multicolumn{2}{c}{$10514,-$}          \\
				\bottomrule[1.0pt]\bottomrule[1.0pt]

				& &\multicolumn{2}{c}{$4^3P_0$}  &  &\multicolumn{2}{c}{$4^3P_2$}  &
				&\multicolumn{2}{c}{$4^1P_1$}   &   &\multicolumn{2}{c}{$4^3P_1$} \\
\end{tabular}}
	\end{center}
\end{table*}

\begin{table*}
	\begin{center}
		\caption{The mass shifts and strong decay width of
each channel for the $6P$-wave bottomonium states, in units of MeV. The caption is the same as that of Table~\ref{4S5S6S}.}
		\label{6P}
		\tabcolsep=0.15cm
		\renewcommand\arraystretch{1.00}
		\scalebox{1.0}{
			\begin{tabular}{cccccccccccccccccccccccc}
				\bottomrule[1.0pt]\bottomrule[1.0pt]
				& &\multicolumn{2}{c}{$6^3P_0$}  &  &\multicolumn{2}{c}{$6^3P_2$}  &
				&\multicolumn{2}{c}{$6^1P_1$}   &   &\multicolumn{2}{c}{$6^3P_1$} \\
				\cline{3-4}\cline{6-7}\cline{9-10}\cline{12-13}
				Channel			&	&$\Delta M$~[11182]  &$\Gamma$~[11207]  &   &$\Delta M$~[11209]  &$\Gamma$~[11216] &
				&$\Delta M$~[11199]  &$\Gamma$~[11205]  &   &$\Delta M$~[11198]  &$\Gamma$~[11203]     \\
				\bottomrule[0.4pt]
				$\bar{B}B$                    &     &$+0.1$                   &$0.1$                     &   &$+0.02$                &$\sim0$             &   &$-$                       &$-$                    &   &$-$                            &$-$                 \\
				$\bar{B}_sB_s$                &     &$\sim0$                  &$\sim0$                   &   &$\sim0$                &$\sim0$             &   &$-$                       &$-$                    &   &$-$                            &$-$                 \\
				$\bar{B}B^*$                  &     &$-$                      &$-$                       &   &$+0.1$                 &$0.03$              &   &$+0.2$                    &$0.2$                  &   &$+0.2$                         &$0.2$                 \\
				$\bar{B}_sB_s^*$              &     &$-$                      &$-$                       &   &$\sim0$                &$0.02$              &   &$\sim0$                   &$0.02$                 &   &$\sim0$                        &$0.03$               \\
				$\bar{B}^*B^*$                &     &$+0.3$                   &$0.5$                     &   &$+0.2$                 &$0.3$               &   &$+0.2$                    &$0.3$                  &   &$+0.2$                         &$0.2$                 \\
\end{tabular}}
	\end{center}
\end{table*}

\begin{table*}
	\begin{center}
		\caption{
The mass shifts and strong decay width of
each channel for the $3D$-, $4D$- and $5D$-wave bottomonium states, in units of MeV. The caption is the same as that of Table~\ref{4S5S6S}.}
		\label{3D4D5D}
		\tabcolsep=0.05cm
		\renewcommand\arraystretch{1.0}
		\scalebox{1.0}{
			\begin{tabular}{cccccccccccccccccccccccc}
				\bottomrule[1.0pt]\bottomrule[1.0pt]
				        & &\multicolumn{2}{c}{$3^3D_1$}  &  &\multicolumn{2}{c}{$3^3D_3$}  &
				&\multicolumn{2}{c}{$3^1D_2$}   &   &\multicolumn{2}{c}{$3^3D_2$} \\
				\cline{3-4}\cline{6-7}\cline{9-10}\cline{12-13}
				Channel			&	&$\Delta M$~[10688]  &$\Gamma$~[10670]  &   &$\Delta M$~[10702]  &$\Gamma$~[10650/10718] &
				&$\Delta M$~[10697]  &$\Gamma$~[10691]  &   &$\Delta M$~[10696]  &$\Gamma$~[10685]     \\
				\bottomrule[0.4pt]
				$\bar{B}B$                    &     &$-0.8$               &$2.1$        &   &$-0.8/+2.5$            &$8.6/4.0$            &   &$-$                &$-$         &   &$-$               &$-$              \\
				$\bar{B}_sB_s$                &     &$\mathbf{-1.0}$      &$-$          &   &$\mathbf{-0.5/-0.9}$   &$-$      /$-$        &   &$-$                &$-$         &   &$-$               &$-$              \\
				$\bar{B}B^*$                  &     &$+8.0$               &$8.6$        &   &$-9.7/+2.7$            &$10.6/25.3$          &   &$-0.2$             &$28.2$      &   &$+4.5$            &$20.7$            \\
				$\bar{B}_sB_s^*$              &     &$-$                  &$-$          &   &$\mathbf{-0.6/-0.9}$   &$-$      /$-$        &   &$\mathbf{-1.9}$    &$-$         &   &$\mathbf{-2.0}$   &$-$               \\
				$\bar{B}^*B^*$                &     &$-23.7$              &$20.0$       &   &$-40.4/+13.0$          &$-$      $/33.6$     &   &$-4.2$             &$52.5$      &   &$-13.7$           &$32.5$            \\
				\emph{Total}                  &     &$-17.5$              &$30.7$       &   &$-52.0/+16.4$          &$19.2/62.9$          &   &$-6.3$             &$80.7$      &   &$-11.2$           &$53.2$            \\
\end{tabular}}
	\end{center}
\end{table*}

\begin{table*}
	\begin{center}
		\caption{The mass shifts and strong decay width of
each channel for the $6D$-wave bottomonium states, in units of MeV. The caption is the same as that of Table~\ref{4S5S6S}.}
		\label{6D}
		\tabcolsep=0.15cm
		\renewcommand\arraystretch{1.0}
		\scalebox{1.0}{
			\begin{tabular}{cccccccccccccccccccccccc}
				\bottomrule[1.0pt]\bottomrule[1.0pt]
				& &\multicolumn{2}{c}{$6^3D_1$}  &  &\multicolumn{2}{c}{$6^3D_3$}  &
				&\multicolumn{2}{c}{$6^1D_2$}   &   &\multicolumn{2}{c}{$6^3D_2$} \\
				\cline{3-4}\cline{6-7}\cline{9-10}\cline{12-13}
				Channel			&	&$\Delta M$~[11308]  &$\Gamma$~[11302]  &   &$\Delta M$~[11323]  &$\Gamma$~[11312] &
				&$\Delta M$~[11318]  &$\Gamma$~[11311]  &   &$\Delta M$~[11317]  &$\Gamma$~[11312]     \\
				\bottomrule[0.4pt]
				$\bar{B}B$                    &     &$+0.02 $              &$0.01 $          &     &$\sim0   $            &$\sim0$           &     &$-$                &$-$              &     &$-$                   &$-$               \\
				$\bar{B}_sB_s$                &     &$\sim0 $              &$0.01 $          &     &$\sim0   $            &$\sim0$           &     &$-$                &$-$              &     &$-$                   &$-$               \\
				$\bar{B}B^*$                  &     &$+0.03 $              &$0.02  $         &     &$\sim0   $            &$\sim0$           &     &$+0.05 $           &$0.01   $        &     &$+0.06    $           &$0.02  $           \\
				$\bar{B}_sB_s^*$              &     &$\sim0 $              &$0.01 $          &     &$\sim0   $            &$\sim0$           &     &$\sim0 $           &$0.03   $        &     &$\sim0    $           &$0.03  $           \\
				$\bar{B}^*B^*$                &     &$+0.1  $              &$0.06  $         &     &$+0.1    $            &$0.1  $           &     &$+0.08 $           &$0.05   $        &     &$+0.07    $           &$0.03  $           \\
				$\bar{B}_s^*B_s^*$            &     &$\sim0 $              &$\sim0$          &     &$\sim0   $            &$\sim0$           &     &$\sim0 $           &$\sim0 $         &     &$\sim0   $            &$0.02  $            \\
\end{tabular}}
	\end{center}
\end{table*}

\begin{table}[htbp]
	\renewcommand\arraystretch{1.2}
	\tabcolsep=0.30cm
	\begin{center}
		\caption{The $\Upsilon(nS)-\Upsilon_1(mD)$ mixing coefficients $c_{S}$ and $c_{D}$, as well as the mixed masses (in MeV), within the hadron loop induced $S$-$D$ mixing framework.}
		\label{SD mixing angle(hadron loop)}
		\scalebox{1.0}{
			\begin{tabular}{cccccccccccccccc}
				\hline\hline
				&$a_{S}$       &$a_{D}$         &$\mathrm{Mass}$   &$\phi_{nS/mD}$             \\	\hline
				$\Upsilon'(4S)$                  &$-0.9999$          &$0.0139$            &$10579$          &$-0.79^{\circ}$    \\
				$\Upsilon'_{1}(3D)$              &$-0.2770$          &$-0.9609$           &$10676$          &$-16.01^{\circ}$     \\[1.0ex]
				
				$\Upsilon'(5S)$                  &$-0.9999$          &$0.0050$            &$10854$          &$-0.29^{\circ}$    \\
				$\Upsilon'_{1}(4D)$              &$-0.0672$          &$-0.9977$           &$10907$          &$-3.85^{\circ}$     \\[1.0ex]
				
				$\Upsilon'(6S)$                  &$-0.9995$          &$0.0310$            &$11015$          &$-1.78^{\circ}$    \\
				$\Upsilon'_{1}(5D)$              &$-0.3108$          &$-0.9505$           &$11104$          &$-18.11^{\circ}$     \\[1.0ex]
				
				$\Upsilon'(7S)$                  &$-0.9933$          &$0.1152$           &$11258$         &$-6.62^{\circ}$    \\
				$\Upsilon'_{1}(6D)$              &$-0.2408$          &$-0.9706$          &$11304$         &$-13.93^{\circ}$     \\
				\hline
				\hline
		\end{tabular}}
	\end{center}
\end{table}

\begin{table*}
	\begin{center}
		\caption{Our predicted dielectron widths (in units of keV) for the vector $b\bar{b}$ states compared with the data and predictions from other literature. Case A stand for the unquenched quark model results predicted with Eqs.(\ref{GammaeeSDmixing}) and (\ref{GammaeeSDmixing2}), where the corrections from both the $S$-$D$ mixing and the continuum components due to coupled-channel effects are included.
Case B stands for the results neglecting the corrections from the $S$-$D$ mixing. Case C stands for the quenched quark model results predicted with Eqs.(\ref{Gammaee}) and (\ref{Gammaee2}).}
		\label{ee Decay Width and Ratio}
		\renewcommand\arraystretch{1.2}
		\tabcolsep=0.12cm
		\scalebox{1.0}{
			\begin{tabular}{ccccccccccccccccccccccccccccccccccccc}
				\bottomrule[1.0pt]\bottomrule[1.0pt]
				State         &BBD~\cite{Badalian:2008ik}   &GVGV~\cite{Gonzalez:2003gx}  &LC~\cite{Li:2009nr}  &GM~\cite{Godfrey:2015dia}  &SOEF~\cite{Segovia:2016xqb}  &WSLM~\cite{Wang:2018rjg}   &Exp~\cite{ParticleDataGroup:2024cfk}    &Case A   &Case B   &Case C \\
				\bottomrule[0.5pt]
				$\Upsilon(1S)$     &$1.317$          &$0.980$              &$1.600$            &$1.440$                  &$0.710$                  &$1.650$                 &$1.340\pm0.018$                     &$\mathbf{1.344}$                 &$1.344$        &$1.344$             \\
				$\Upsilon(2S)$     &$0.614$          &$0.410$              &$0.640$            &$0.730$                  &$0.370$                  &$0.821$                 &$0.612\pm0.011$                     &$\mathbf{0.557}$                 &$0.557$        &$0.557$             \\
				$\Upsilon(3S)$     &$0.448$          &$0.270$              &$0.440$            &$0.530$                  &$0.270$                  &$0.569$                 &$0.443\pm0.008$                     &$\mathbf{0.408}$                 &$0.408$        &$0.408$             \\
				$\Upsilon(10580)$  &$0.370$          &$0.200$              &$0.350$            &$0.390$                  &$0.210$                  &$0.431$                 &$0.272\pm0.029$                     &$\mathbf{0.209}$                 &$0.209$        &$0.343$             \\
				$\Upsilon(10860)$  &$0.316$          &$0.160$              &$0.290$            &$0.330$                  &$0.180$                  &$0.348$                 &$0.220\pm0.120$~\cite{CLEO:1984vfn} &$\mathbf{0.224}$                 &$0.224$        &$0.298$            \\
				$\Upsilon(11020)$  &$0.274$          &$0.120$              &$0.250$            &$0.270$                  &$0.150$                  &$0.286$                 &$0.130\pm0.030$                     &$\mathbf{0.100}$                 &$0.101$        &$0.273$             \\
				$\Upsilon(7S)$     &$...$            &$0.100$              &$0.220$            &$...$                    &$...$                    &$0.243$                 &$...$                               &$\mathbf{0.130}$                 &$0.137$        &$0.250$              \\
				$\Upsilon_1(1D)$   &$0.62$E$-3$      &$3.70$E$-4$          &$...$              &$1.38$E$-3$              &$1.40$E$-3$              &$1.88$E$-3$             &$...$                               &$\mathbf{1.08}$E$\mathbf{-3}$    &$1.08$E$-3$    &$1.08$E$-3$           \\
				$\Upsilon_1(2D)$   &$1.08$E$-3$      &$5.80$E$-4$          &$...$              &$1.99$E$-3$              &$2.50$E$-3$              &$2.81$E$-3$             &$...$                               &$\mathbf{2.13}$E$\mathbf{-3}$    &$2.13$E$-3$    &$2.13$E$-3$           \\
				$\Upsilon_1(3D)$   &$1.44$E$-3$      &$6.70$E$-4$          &$...$              &$2.38$E$-3$              &$...$                    &$3.00$E$-3$             &$...$                               &$\mathbf{0.028}$                 &$1.95$E$-3$    &$3.16$E$-3$           \\
				$\Upsilon_1(4D)$   &$1.71$E$-3$      &$7.90$E$-4$          &$...$              &$2.18$E$-3$              &$...$                    &$3.00$E$-3$             &$...$                               &$\mathbf{7.19}$E$\mathbf{-3}$    &$2.80$E$-3$    &$4.11$E$-3$           \\
				$\Upsilon_1(5D)$   &$1.90$E$-3$      &$7.10$E$-4$          &$...$              &$...$                    &$...$                    &$3.02$E$-3$             &$...$                               &$\mathbf{0.019}$                 &$1.66$E$-3$    &$5.03$E$-3$            \\
				$\Upsilon_1(6D)$   &$...$            &$...$                &$...$              &$...$                    &$...$                    &$...$                   &$...$                               &$\mathbf{0.023}$                 &$4.14$E$-3$    &$5.92$E$-3$           \\
				\bottomrule[1.0pt]\bottomrule[1.0pt]
		\end{tabular}}
	\end{center}
\end{table*}

\section*{Acknowledgements }
We thank useful discussions from Qi Li and Yu Lu.
This work is supported by the National Natural Science Foundation of China under Grants Nos.12175065 and 12235018 (X.H.Z), 12175239 and 12221005 (J.J.W), and by the National Key R $\&$ D Program of China under Contract No. 2020YFA0406400 (J.J.W), and supported by Chinese Academy of Sciences under Grant No. YSBR-101 (J.J.W).

\begin{appendix}

\section{Mass shift and spectral density function}

\begin{figure*}
	\centering \epsfxsize=13.0 cm \epsfbox{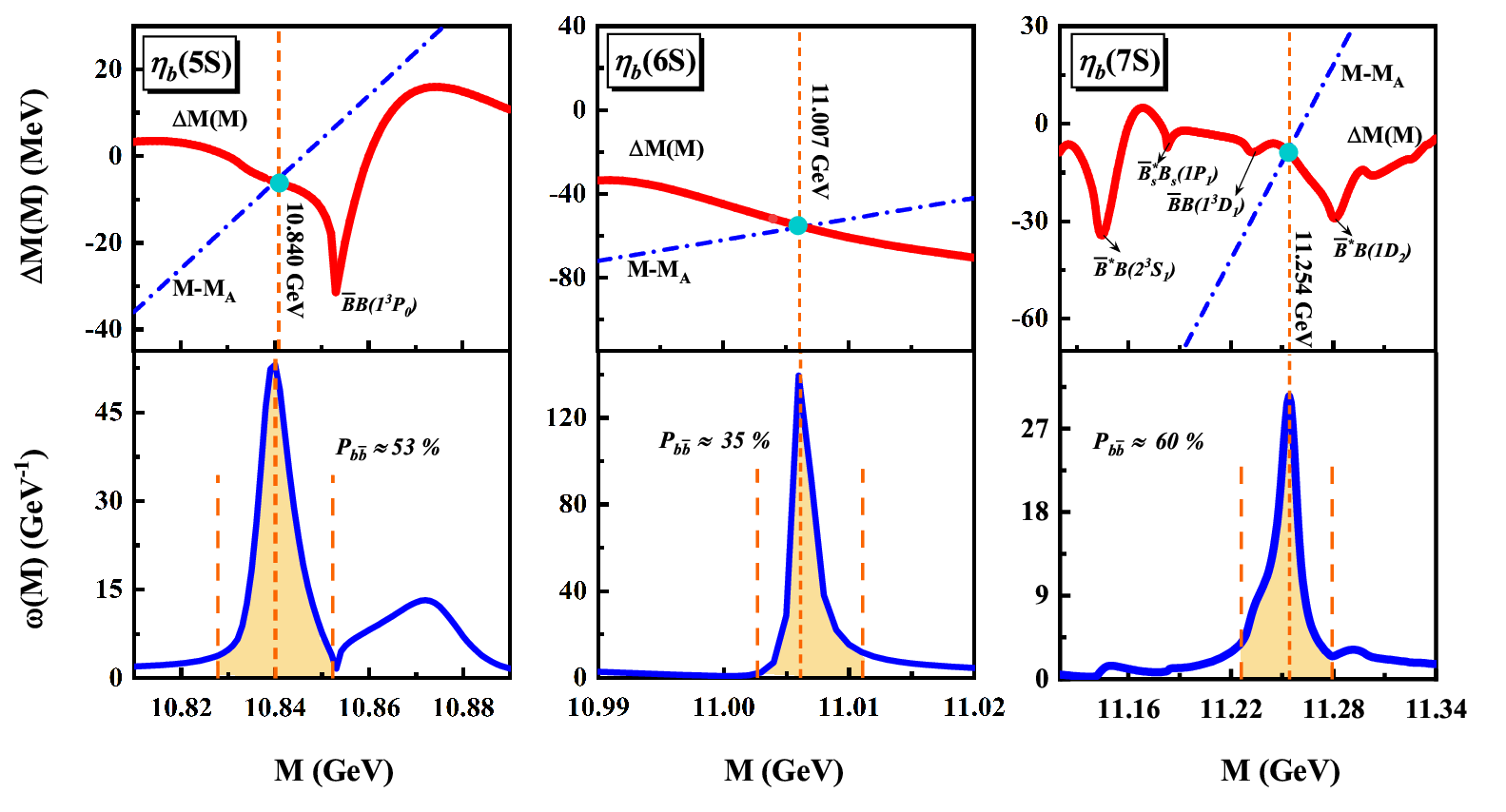} \vspace{-0.2 cm} \caption{The line shapes of the mass shift function $\Delta M(M)$ (upper panel) and spectral density function $\omega(M)$ (lower panel) for the $\eta_{b}(5S,6S,7S)$ states. The solution of the coupled-channel Eq.(\ref{M=MA+Delta M}) corresponds to the intersection point between two lines $\Delta M(M)$ and $M - M_A$ in the upper panel.
Some important coupled-channels are also marked in the upper panel. The shaded area in
the spectral density function is the integral interval for estimating the $b\bar{b}$ component. }\label{bbbar1S0MSandSF}
\end{figure*}

\begin{figure*}
	\centering \epsfxsize=17.4 cm \epsfbox{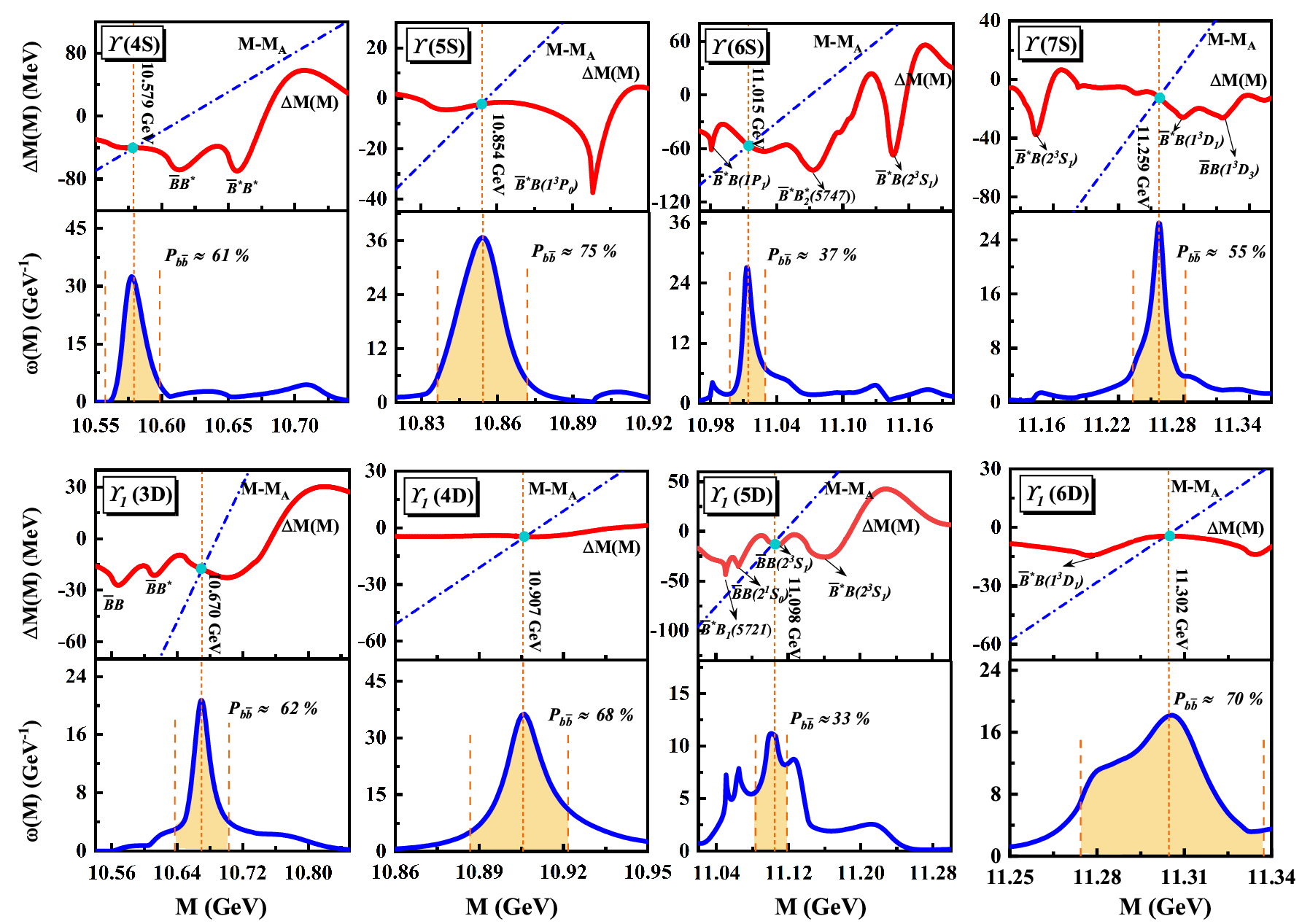} \vspace{-0.0 cm} \caption{
The line shapes of the mass shift function $\Delta M(M)$ (upper panel) and spectral density function $\omega(M)$ (lower panel) for the vector $b\bar{b}$ states. The caption is the same as that of Fig.~\ref{bbbar1S0MSandSF}.}\label{bbbarVectorMSandSF}
\end{figure*}

\begin{figure*}
\epsfxsize=17.8 cm \epsfbox{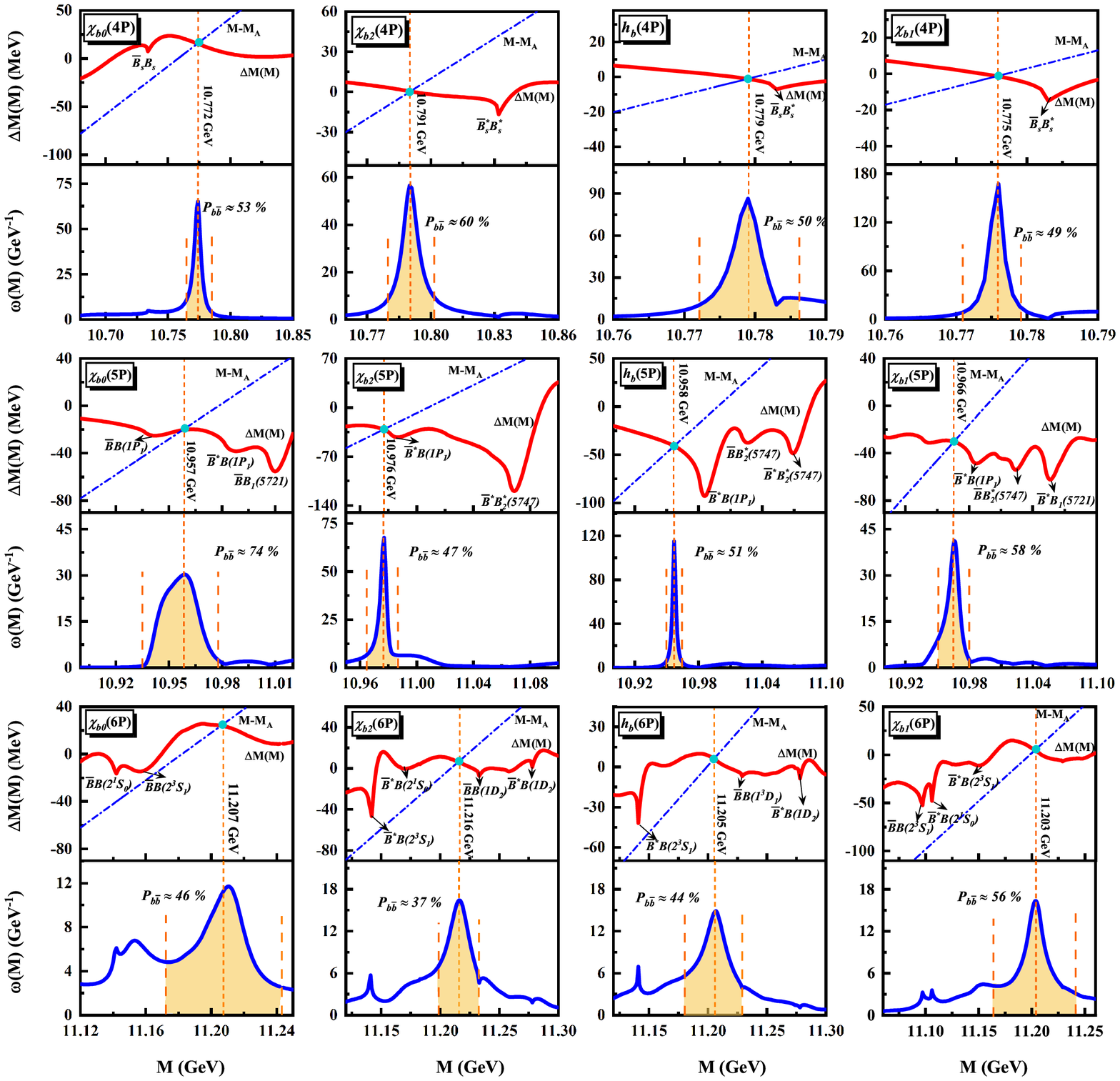}\vspace{-0.0 cm} \caption{
The line shapes of the mass shift function $\Delta M(M)$ (upper panel) and spectral density function $\omega(M)$ (lower panel) for the $4P$-, $5P$-, and $6P$-wave states. The caption is the same as that of Fig.~\ref{bbbar1S0MSandSF}.}
\label{bbbarPwaveMSandSF}
\end{figure*}

\begin{figure*}
\epsfxsize=13.0 cm \epsfbox{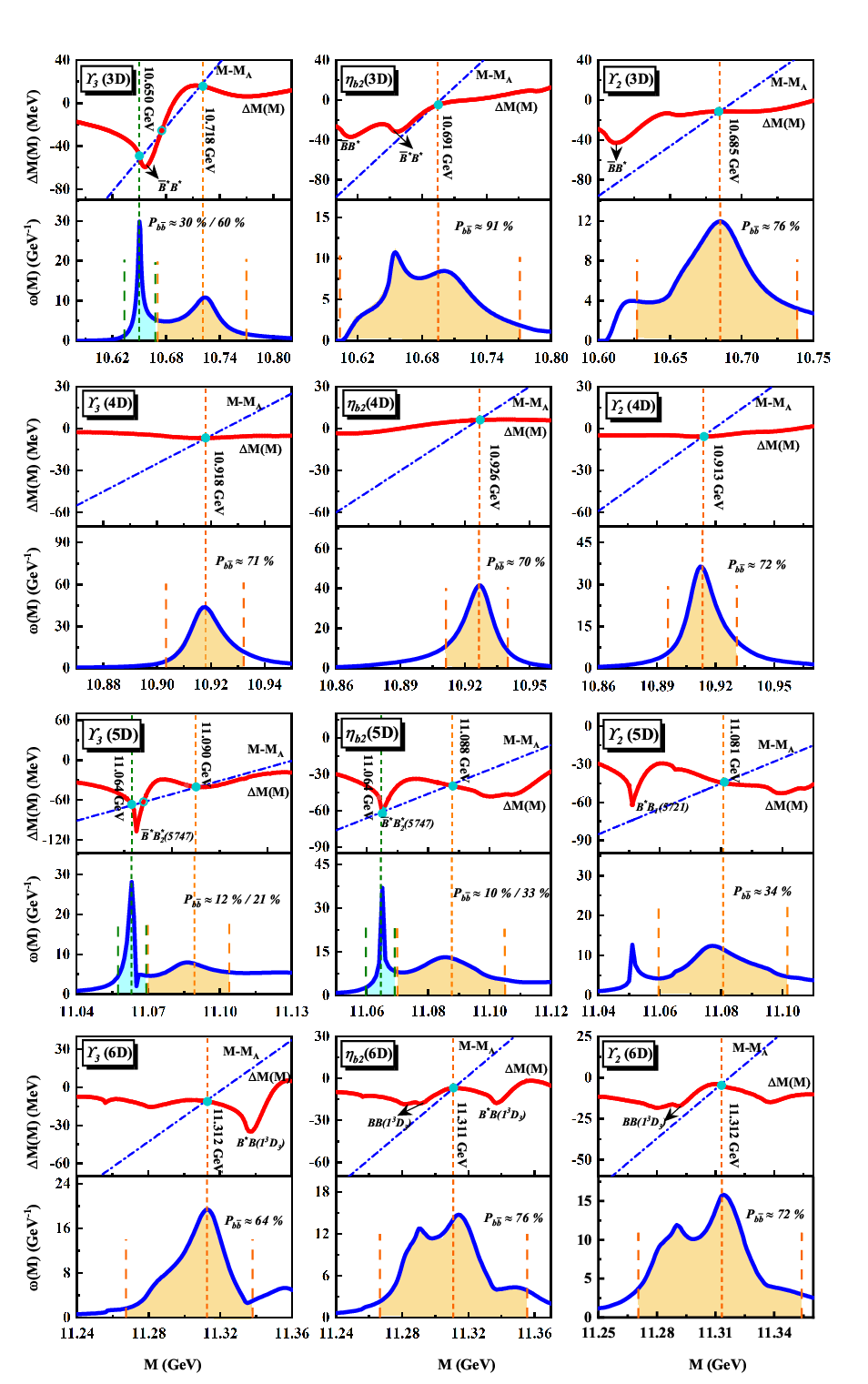}\vspace{-0.0 cm} \caption{The line shapes of the mass shift function $\Delta M(M)$ (upper panel) and spectral density function $\omega(M)$ (lower panel) for the $\Upsilon_{2,3}(nD)$ and $\eta_{b2}(nD)$ ($n=3,4,5,6$) states. The caption is the same as that of Fig.~\ref{bbbar1S0MSandSF}.}
\label{bbbarOthersDwaveMSandSF}
\end{figure*}

In this appendix, we present the total mass shifts as a function of the physical mass, along with the spectral density function for the $b\bar{b}$ states, as shown in Figs.~(\ref{bbbar1S0MSandSF})-(\ref{bbbarOthersDwaveMSandSF}).

\end{appendix}

\bibliographystyle{unsrt}

\end{document}